\PassOptionsToPackage{numbers,sort&compress}{natbib}
\documentclass[preprintnumbers,amsmath,amssymb,prd, notitlepage,nofootinbib, superscriptaddress]{revtex4}

\usepackage{amsfonts,amssymb,amsmath}
\usepackage{gensymb}
\usepackage{color}
\usepackage{graphicx}
\usepackage{multirow}
\usepackage[utf8]{inputenc}
\usepackage{aas_macros}
\usepackage{hyperref}

\usepackage[normalem]{ulem}

\newcommand{\greaterthanapprox}{\mathrel{\vcenter{
  \offinterlineskip\halign{\hfil$##$\cr
    >\cr\noalign{\kern2pt}\sim\cr\noalign{\kern-2pt}}}}}
    
    \newcommand{\lessthanapprox}{\mathrel{\vcenter{
  \offinterlineskip\halign{\hfil$##$\cr
    <\cr\noalign{\kern2pt}\sim\cr\noalign{\kern-2pt}}}}}
    
\newcommand{\lb}{\left(}
\newcommand{\rb}{\right)}

\newcommand{\Planck}{{\it Planck}~}

\newcommand{\tcb}{\textcolor{blue}}

\newcommand{\be}{\begin{equation}}        
\newcommand{\ee}{\end{equation}}

\newcommand{\bestfitTstandard}{10.14}
\newcommand{\bestfitbetastandard}{1.77}
\newcommand{\bestfitTplus}{11.95}
\newcommand{\bestfitbetaplus}{1.59}

\newcommand{\NPIPE}{\texttt{NPIPE}}

\begin{document}

\title{Component-separated, CIB-cleaned  thermal Sunyaev--Zel'dovich maps from \textit{Planck} PR4 data with a flexible public needlet ILC pipeline}

\author{Fiona McCarthy}
\email{fmccarthy@flatironinstitute.org}

\affiliation{Center for Computational Astrophysics, Flatiron Institute, New York, NY, USA 10010}

\author{J.~Colin Hill}
\email{jch2200@columbia.edu}

\affiliation{Department of Physics, Columbia University, 538 West 120th Street, New York, NY, USA 10027}
\affiliation{Center for Computational Astrophysics, Flatiron Institute, New York, NY, USA 10010}

\date{\today}

\begin{abstract}

We use the full-mission \textit{Planck} PR4 data to 
construct maps of the thermal Sunyaev--Zel'dovich effect (Compton-$y$ parameter) in our Universe.  To do so, we implement a custom needlet internal linear combination (NILC) pipeline in a Python package, \texttt{pyilc}, which we make publicly available.  We publicly release our Compton-$y$ maps, which we construct using various constrained ILC (``deprojection'') options in order to minimize contamination from the cosmic infrared background (CIB) in the reconstructed signal.  In particular, we use a moment-based deprojection which minimizes sensitivity to the assumed frequency dependence of the CIB.  Our code \texttt{pyilc} performs needlet or harmonic ILC on mm-wave sky maps in a flexible manner, with options to deproject various components on all or some scales. We validate our maps and compare them to the official \textit{Planck} 2015 $y$-map, finding that we obtain consistent results on large scales and 10-20\% lower noise on small scales.  We expect that these maps will be useful for many auto- and cross-correlation analyses; in a companion paper, we use them to measure the tSZ -- CMB lensing cross-correlation.  We anticipate that \texttt{pyilc} will be useful both for data analysis and for pipeline validation on simulations to understand the propagation of foreground components through a full NILC pipeline. 
\end{abstract}
 
\maketitle

\section{Introduction}

Observations of the millimeter-wave sky contain contributions from many sources, both Galactic and extragalactic. The signals of cosmological interest include, of course, the primary cosmic microwave background (CMB)~\cite{1965ApJ...142..414D,1965ApJ...142..419P} --- the photons that have been free-streaming since recombination without scattering --- along with the Sunyaev--Zel'dovich (SZ) effect~\cite{1969Ap&SS...4..301Z,1970Ap&SS...7....3S}, which is sourced by the scattering of the CMB photons from free electrons in the late Universe, and the cosmic infrared background (CIB)~\cite{1996A&A...308L...5P,1998ApJ...508..106D,1998ApJ...508..123F}, which is the thermal radiation of dust grains in star-forming galaxies that are heated by starlight. As we can extract very different information from the different signals, it is useful to be able to separate them. It is common to do so by observing the sky at multiple frequencies and separating the signals based on their differing frequency behavior. In particular, the CMB intensity behaves as a perfect blackbody, with a temperature of 2.726 $\mathrm{K}$~\cite{1996ApJ...473..576F,1999ApJ...512..511M,2009ApJ...707..916F}. The thermal SZ (tSZ) effect --- the scattering of CMB photons by high-temperature electrons --- induces a well-understood distortion in this spectrum, allowing the tSZ anisotropies to be separated cleanly from the primary CMB. 

Many component separation algorithms exist for performing such separation (see, e.g.,~\cite{2003ApJS..148...97B,2003MNRAS.345.1101M,2008ApJ...676...10E,2017arXiv171000621S,2009A&A...493..835D,2011MNRAS.418..467R,2008ISTSP...2..735C,2013A&A...558A.118H}). In particular, we focus on the internal linear combination (ILC), a method that creates linear combinations of maps using the well-understood frequency {behavior} of a signal of interest, such as the blackbody CMB or the tSZ effect, and which has been applied to CMB data for decades~\cite{2003ApJS..148...97B}. Importantly, this is a ``blind'' component separation technique, in that the frequency behavior of the contaminants does not need to be known a priori. This is necessary, as the CIB --- while observed to generally behave as a modified blackbody, with its emission dominant (with respect to the CMB) at higher frequencies ($\gtrsim$353 GHz, although on small scales the CIB anisotropies are dominant at lower frequencies than this) --- does not have a well-understood spectral energy distribution (SED){, or frequency dependence,} that can be derived from first principles. However, when the SED of a contaminant is known, this information can be used in the ILC to create a map that is completely insensitive to this contaminant~\cite{2009ApJ...694..222C,2011MNRAS.410.2481R} (or partially less sensitive than the unconstrained estimate~\cite{2021PhRvD.103j3510A}), although at the expense of higher variance in the overall reconstruction.  This ``deprojection'' of foreground components, known as a ``constrained'' ILC, is useful for example in the case of cross-correlations, when one is cross-correlating a component of interest with another signal that is highly correlated with a contaminant.

In this work, we use a needlet ILC (NILC) algorithm to construct maps of the tSZ effect using \textit{Planck} PR4 (\NPIPE) data~\cite{2020A&A...643A..42P}; we note that several $y$-maps have previously been made with \textit{Planck} data~\cite{2014JCAP...02..030H,2016A&A...594A..22P,2022MNRAS.509..300T,2023arXiv230510193C}. We also release maps in which we have deprojected the CIB in various ways, both with a standard CIB deprojection using an estimate of the CIB SED, and in a more theory-independent method where we deproject the first moments of the CIB SED, following the methods of~\cite{2017MNRAS.472.1195C}; in a companion paper~\cite{Paper2} we use these maps to measure the cross-correlation of the tSZ signal with CMB lensing convergence ($\kappa$) in \textit{Planck} PR4 data using a publicly-available \textit{Planck} $\kappa$ reconstruction~\cite{2022JCAP...09..039C}. 

We validate our Compton-$y$ maps  and compare them to that from the 2015 \textit{Planck} release~\cite{2016A&A...594A..22P}. While this work was in the final stages of preparation, a similar analysis performing NILC on \textit{Planck} PR4 data appeared~\cite{2023arXiv230510193C}. We defer a comparison to this map to future work.

A central purpose of this paper is also to release a user-friendly Python package \texttt{pyilc}, a flexible code which can perform needlet and harmonic ILC on curved-sky maps for various specifications of needlet domains. This is (to our knowledge) the first publicly available needlet ILC code.

This paper is structured as follows.  In Section~\ref{sec:data} we describe the datasets we use to construct the ILC maps. Section~\ref{app:nilc_description} presents a pedagogical overview of the NILC algorithm and the ``ILC bias''.  In Section~\ref{sec:ilc_ymap} we describe our pipeline to construct a $y$-map from the data using the NILC approach, including our exact analysis settings. Section~\ref{sec:deprojection_choices} discusses the various CIB deprojection options that we implement.  In Section~\ref{sec:comp_planck} we validate our $y$-maps, compute their power spectra, compare them to that of the official \textit{Planck} release, and quantify the amount of CIB contamination in the various deprojections.   We discuss our results and conclude in Section~\ref{sec:conclusion}.

\section{Data}\label{sec:data}

We use the single-frequency full-mission maps from the \textit{Planck} \NPIPE~data release (PR4)~\cite{2020A&A...643A..42P} in our analysis. These maps are provided in $\mathrm{\mu K}$ (CMB thermodynamic temperature units) at all frequencies, so we do not need to convert between $\mathrm{Jy/sr}$ and $\mathrm{\mu K}$ (as might have been necessary for previous releases of the \textit{Planck} single-frequency maps, some of which provided the highest-frequency maps in  $\mathrm{Jy/sr}$). Note that the kinematic solar dipole, which is sourced by the Doppler boosting of the CMB monopole due to our proper motion with respect to the CMB rest frame (in particular our motion through the Galaxy), is not subtracted from these maps. As this is much brighter than the intrinsic CMB fluctuations (with an amplitude of $\approx 3367 \, \mathrm{\mu K}$~\cite{2020A&A...643A..42P}, cf.~the characteristic CMB fluctuation amplitude of $\sim 40 \, \mathrm{\mu K}$), we subtract it from the maps before using them. To ensure we subtract the same dipole from each map, we use the solar dipole estimation from the \texttt{Commander} component-separation analysis of the \NPIPE~maps\footnote{This is available on NERSC at \texttt{\$CFS/cmb/data/planck2020/all\_data/commander\_dipole\_templates/planck/dipole\_CMB\_n4096\_K.fits}}. We also subtract the mean of each map before further analysis.

We use the 30, 44, 70, 100, 143, 217, 353, 545, and 857 GHz single-frequency maps. The observed sky signal in these maps is convolved with the \textit{Planck} instrument beam at the corresponding frequency. We approximate these beams as being Gaussian, with full width at half maximum values (FWHMs) given in Table~\ref{tab:fwhm}~\cite{2020A&A...641A...1P}. To characterize fully the frequency response of each map, we use the passbands given in Refs.~\cite{2014A&A...571A...2P,2014A&A...571A...9P}.

\begin{table*}
\begin{tabular}{|c||c|c|c|}\hline
Frequency (GHz) & Beam FWHM (arcmin) & Noise ($\mathrm{\mu K}$ arcmin) & Noise power spectrum amplitude ($\mathrm{\mu K}^2$)\\\hline\hline
30 & 32.29& 150& 0.00190\\\hline
44 & 27.94& 162&0.00222 \\\hline
70 & 13.08&210 & 0.00373\\\hline
100 & 9.66& 77.4&0.000507 \\\hline
143 & 7.22&33 & 9.21$\times10^{-5}$\\\hline
217 & 4.90&46.8& 0.000185\\\hline
353 & 4.92&154 &0.00200 \\\hline
545 & 4.67&806.7 &0.0551\\\hline
857 & 4.22 &19115 &30.9\\\hline
\end{tabular}
\caption{Characterizing features of the \textit{Planck} experiment, in particular the beam FWHM and approximate white noise levels for each frequency channel. We quote the noise both in $\mathrm{\mu K}$ arcmin and in $\mathrm{\mu K}^2$; the latter is calculated from the former by converting to $\mu \mathrm{K}$ radians (i.e., multiplying by $\pi/(180\times60)$) and then squaring. This information comes from Table 4 of~\cite{2020A&A...641A...1P}. Note that we have applied a $\mathrm{Jy/sr}$-to-$\mathrm{\mu K}$ conversion factor to the values quoted for 545 and 857 GHz, as we analyze the maps in $\mathrm{\mu K}$. }
\label{tab:fwhm}
\end{table*}

We also use the half-ring split maps from the \texttt{NPIPE} release. These maps are subset maps of the full-mission maps described above, each with the same passbands, beams, etc.~as the full-mission maps described above, but with independent noise realizations; as such, each half-ring map is noisier than the full-mission map, but they are useful for power spectrum analysis as their cross-power spectrum does not contain bias due to correlations in the instrumental noise. With these maps, we build two independent split maps of the Compton-$y$ signal.

We use a needlet ILC algorithm to construct a full-sky tSZ map from these frequency maps, with various choices for contaminant deprojection. We discuss the specific details of our needlet ILC pipeline in detail in Section~\ref{sec:ilc_ymap}.

\section{The needlet ILC algorithm}\label{app:nilc_description}

In this section, we discuss the ILC and NILC in general, beginning in Part~\ref{sec:ILC_general} with a general definition of ILC, and describing the NILC in Part~\ref{sec:nILC_general}. We discuss the constrained ILC, which can be used to deproject specific foreground components with known frequency dependence, in Part~\ref{sec:constrained_ILC}. Finally in Section~\ref{sec:ILC_bias}, we discuss the ILC bias. Note that in this section, we remain very general, specializing to the specific details of our NILC pipeline in Section~\ref{sec:ilc_ymap}.

 Throughout, we use lower-case Latin indices $i,j,k,l,m,n,...$ to label components of vectors or matrices in frequency space; subscript $\ell, m$ indices refer to discrete harmonic-space multipole coefficients (while the symbol $m$ may seem overloaded, it will always be clear from context to what we refer). Greek indices $\alpha,\beta,...$ refer to components in the space of sky components that we preserve or deproject. 
\subsection{ILC}\label{sec:ILC_general}

The internal linear combination (ILC) (see, e.g.,~\cite{2003ApJS..148...97B} for an early application reconstructing a CMB map using the \textit{WMAP} data) relies on knowledge of the frequency dependence of a signal to isolate it by taking a linear combination of multi-frequency measurements. In particular, assuming the temperature anisotropy $T$ in frequency channel $i$ in direction $\hat{n}$ is given by
\be
T_i(\hat n) = a_i s(\hat n) + n_i(\hat n),
\ee
where $s(\hat n)$ is the signal of interest, $a_i$ is its \textit{known} frequency dependence or spectral energy distribution (SED), and $n_i(\hat n)$ is all other sources of intensity, which includes atmospheric or instrumental noise as well as cosmological or Galactic foregrounds (or any other signal that the detector measures). Any linear combination of the temperature maps
\be
\tilde s (\hat n ) = \sum _i w_i T_i (\hat n)
\label{eq.linearcomb}
\ee
that obeys the condition
\be
\sum_i w_i a_i = 1
\label{ILC_condition}
\ee
is unbiased to the signal of interest, in that it can be written as
\be
\tilde s(\hat n ) = s(\hat n) + \tilde n (\hat n)
\ee
where $\tilde n (\hat n)$ is uncorrelated with the signal $s(\hat n)$ (provided the noise and the foregrounds are uncorrelated with the signal --- note that this assumption is in fact broken for the case of the tSZ effect, which is correlated with the cosmic infrared background and other LSS-induced foregrounds). The weights $w_i$ in Eq.~\eqref{eq.linearcomb} that result in a minimum-variance estimate of the signal are 
\be
\mathbf{w} = (\mathbf{a}^T \mathbf{\mathcal{C}}^{-1} \mathbf{a})^{-1} \mathbf{a}^T \mathbf{\mathcal{C}}^{-1} ,
\label{ILC_weights_app}
\ee
where $\mathbf{\mathcal{C}}$ is the covariance matrix of the data. These weights can be found straightforwardly by a minimization with the method of Lagrange multipliers to preserve the constraint~(e.g.,~\cite{Eriksen:2004jg}). In explicit index notation, this can be written 
\be
w_i = \frac{a_j \left(\mathcal{C}^{-1}\right)_{ij} }{a_k \left(\mathcal{C}^{-1}\right)_{kl}a_l  },
\ee
where, here and to follow, the Einstein summation convention is assumed, i.e., repeated indices are summed over. 

In the ILC approach, the frequency-frequency covariance matrix is estimated directly from the data, and so the only ``external'' knowledge that is needed is the knowledge of the signal's frequency dependence $a_i$ (hence the nomenclature ``internal'').  The covariance can be measured and the weights applied in various bases: for example, in real space, by measuring $\mathcal{C}_{ij}$ over the entire maps (or subregions of the maps) and calculating one weight for each frequency map; or in harmonic space, by calculating an $\ell$-dependent covariance matrix $\mathcal{C}^{ij}_\ell$ and applying $\ell$-dependent weights to the harmonic coefficients of the frequency maps in separate multipole bins. The former is ideal if there is no scale dependence to the noise properties, e.g., if the noise and the signal all have power spectra with similar $\ell$-dependence. If the foregrounds have different $\ell$-dependence to the signal, or if different foregrounds are relevant at different scales in the different frequency channels, or if the instrumental noise becomes dominant at different values of $\ell$, then a harmonic ILC is more appropriate. We describe briefly each of these domains below.

\subsubsection{Real-space ILC}

In a real-space domain $\mathcal {D}^{\mathrm{real}}$, the temperature in frequency band $i$ on the sphere $T_i(\Omega)$ is defined on a discrete basis of pixels $p(\hat n)$: 
\be
T_i(\Omega) = \sum _{\hat n} T_i(\hat n) p(\hat n).
\ee
In general, the pixelized coefficients $T_i(\hat n)$ can be found by
\be
T_i(\hat n) = \int d\Omega T_i(\Omega) p(\hat n);
\ee
in the most simple version, $p(\hat n)$ is a step function with unit value within the area defined by the pixel and zero elsewhere.

In this domain, the frequency-space covariance matrix $\mathcal{C}$ is calculated according to
\begin{align}
\mathcal{C}_{ij} =& \left<\left(T_i(\hat n) - \left< T_i\right>\right)\left(T_j(\hat n)  - \left< T_j\right>\right)\right>\\
 = &\frac{1}{N_{\mathrm{pix}-1}}\sum_{\hat n \in \mathcal {D}^{\mathrm{real}}}\left(T_i(\hat n) - \left< T_i\right>\right)\left(T_j (\hat n) - \left< T_j\right>\right),
\end{align}
where 
\be
\left< T_i\right>\equiv \frac{1}{N_{\mathrm{pix}}}\sum_{\hat n \in \mathcal {D}^{\mathrm{real}}} T_i (\hat n)
\ee 
denotes the mean (over $\mathcal D^{\mathrm{real}}$) of $T_i(\hat n)$, and $N_{\mathrm{pix}}=\sum_{\hat n \in \mathcal {D}^{\mathrm{real}}} p (\hat n)$ denotes the number of pixels in the domain $ \mathcal{D}^{\mathrm{real}}$.

If $ \mathcal{D}^{\mathrm{real}}$ covers the entire sphere (or the entire area of the unmasked map), the ILC weights then only depend on the frequency channel $i$ and are applied uniformly to the entire map. Alternatively, $\mathcal{D}^{\mathrm{real}}$ can be a subset of the map, allowing for spatial dependence of the weights; this may be appropriate for statistically anisotropic fields, such as when dealing with the foreground emission from our own galaxy, which is highly statistically anisotropic. However, it should be noted that, while the true temperature is preserved within the domains, correlations on scales larger than each domain are lost in such an application.

\subsubsection{Harmonic-space ILC}

The basis of the harmonic domain is determined by the spherical harmonic functions $Y_{\ell m}(\hat n)$, which are are the solutions to the Laplace equation on the sphere:
\be
T(\hat n) = \sum _{\ell m} T_{\ell m}Y_{\ell m}(\hat n).
\ee
The spherical harmonic coefficients $T_{\ell m}$ are related to the pixel coefficients $T({\hat n})$ according to
\be
T_{\ell m} = \sum_{\hat n}  T({\hat n}) Y^*_{\ell m}(\hat n)\tcb{,}
\ee
where $Y_{\ell m}^*$ denotes the complex conjugate of $Y_{\ell m}$, and the inverse transformation is 
\be
  T({\hat n}) = \sum_{\ell m} T_{\ell m} Y_{\ell m}(\hat n).
\ee

In this domain, we can calculate the covariance matrix in an $\ell$-dependent manner.  Defining the measured power spectrum $\hat C^{ij}_\ell$ according to 
\be
\hat  C^{i j}_\ell  = \left<T^i_{\ell m} T^j_{\ell m}\right> =\sum _{m} \frac{ T^i_{\ell m} T^{j*}_{\ell m}}{2 \ell+1},
\ee
the covariance matrix at scale $\ell$, $\mathcal C_{ij}(\ell)$, is given by
\be
\mathcal C_{ij}(\ell) = \frac{2\ell+1}{4\pi} \hat C^{i j}_\ell.
\ee
In practice, it can be useful to calculate $\mathcal C$ in a domain  $D^{\mathrm{harm}}(\ell_0)$ defined by a \textit{band} of multipoles, for example centered at some value $\ell_0$ with width $\Delta \ell$. In this case, $\mathcal C_{ij}(\ell_0)$ is given by

\be
\mathcal C_{ij}(\ell_0) =  \sum _{\ell = \ell_0- \Delta \ell/2}^{\ell = \ell_0+ \Delta \ell/2} \frac{2\ell+1}{4\pi}\hat C_\ell^{ij}.
\ee

The $\ell$-dependent weights, calculated from Eq.~\eqref{ILC_weights_app}, can then be applied to the spherical harmonic coefficients $T^i_{\ell m}$ with $\ell \in \mathcal D^{\mathrm{harm}}(\ell_0)$, i.e., with $\ell_0- \Delta \ell/2<\ell<\ell_0+ \Delta \ell/2$. Note that it is not necessary to define disjoint bands (they can overlap, if desired); we could alternatively define a separate such bin at every $\ell$, with $ C_{ij}(\ell_0)$ calculated using information from the surrounding multipoles\footnote{However, currently our harmonic ILC implementation in \texttt{pyilc} only uses disjoint $\ell$-bands.}. 

Multipole-dependent weights are appropriate when there is different scale-dependent behavior of the foregrounds and instrumental noise, including foreground SEDs that depend on $\ell$, to allow for the variance to be adaptively minimized on all scales. However,  the harmonic ILC, while optimal for a statistically isotropic field, is not equipped to deal with statistical anisotropy, such as that from Galactic foreground components.

\subsection{Needlet ILC (NILC)}\label{sec:nILC_general}

To combine the advantages of both the real-space and harmonic-space ILC approaches, the ILC domain can be defined on a needlet frame~\cite{2009A&A...493..835D}. Needlets~\cite{doi:10.1137/040614359} are a construction of a spherical wavelet frame (a frame is similar to an over-complete basis) which allows for simultaneous localization in real and harmonic space. The frame is defined first by a set of harmonic-space window functions  (indexed by capital Latin letters $I,J,\ldots$) $h_\ell^I$ which obey
\be
\label{eq.NILCfilterreq}
\sum _I \left( h _\ell ^I\right)^2 =1
\ee
at each $\ell$. Each $I$ specifies a different ``needlet scale''. The needlets are further defined by real-space domains $\mathcal{D}^{\mathrm{real},I}_{\hat n}$ associated with each needlet scale, where  $\mathcal{D}^{\mathrm{real},I}_{\hat n}$ can be defined independently at each pixel $\hat n$ (and  $\mathcal{D}^{\mathrm{real},I}_{\hat n}$  are not disjoint over the pixels). 

In practice, the needlet ILC consists of the following steps: 
\begin{enumerate}
\item For each needlet scale $I$, each frequency map $i$ is filtered in harmonic space according to the window function $h _\ell ^I$:
\be
T_i^I(\hat n) = \sum _{\ell m} h_\ell^I T^i_{\ell m}Y_{\ell m}(\hat n).\label{needlet_coefficients}
\ee
The maps $T_i^I(\hat n) $ are referred to as the ``needlet coefficients''. Eq.~\eqref{needlet_coefficients} amounts to taking a spherical harmonic transformation of each $T_i(\hat n)$, filtering the coefficients by multiplying them by $h_\ell^I$, and taking the inverse spherical harmonic transformation of the result. 

\item For each pixel $\hat n$ at each needlet scale $I$, the local frequency-frequency covariance matrix $\mathcal C_{ij}(I, \hat n)$ is calculated on a domain $\mathcal {D}^{\mathrm{real},I}_{\hat n}$. This results in a set of $N_I$ matrices of size $N_{\mathrm{freq}}\times N_{\mathrm{freq}}$, at each pixel $\hat n$:
\be
\mathcal C^I _{ij} (\hat n)= \frac{1}{N_{\mathrm{pix}}-1} \sum _{\hat n^\prime \in \mathcal D^{\mathrm{real},I}_{\hat n}} \left(T_i^I(\hat n^\prime)-\left<T_i^I\right>\right)\left(T_j^I(\hat n^\prime)-\left<T_j^I\right>\right)
\ee
where $N_{\mathrm{pix}}$ is the number of pixels in $\mathcal D^{\mathrm{real},I}_{\hat n}$ and $\left<T_i^I\right>$ is the mean of the temperature on $ \mathcal D^{\mathrm{real},I}_{\hat n}$:
\be
\left<T_i^I\right> \equiv  \frac{1}{N_{\mathrm{pix}}} \sum _{\hat n^\prime \in \mathcal D^{\mathrm{real},I}_{\hat n}}T_i^I(\hat n^\prime).
\ee

\item The needlet ILC weights $w_i^I(\hat n)$ are then calculated at each needlet scale $I$ according to Eq.~\eqref{ILC_weights_app}. Note that the inverse of the covariance matrix $\mathcal C^I _{ij} (\hat n)$ is the inverse in the frequency basis, and the inversion is performed separately at each pixel $\hat n$. These weights are then applied to the needlet coefficients to build the ILC estimate at each needlet scale:
\be
T_{\mathrm{ILC}}^I(\hat n) = \sum _i w_i^I(\hat n) T_i^I(\hat n) \,.
\label{eq.ILCperneedlet}
\ee

\item The spherical harmonic coefficients of each ILC estimate are then computed and filtered (again) by the needlet window functions, then transformed back to pixel space:
\be
z^I (\hat n)  = \sum _{\ell m}  h_\ell^I{T_{\mathrm{ILC}, \ell m}^I}Y_{\ell m}(\hat n) \,.
\ee
These final maps are then added to arrive at the final NILC estimate:
\be
T^{\mathrm{NILC}}(\hat n) = \sum _I z^I (\hat n) \,.
\label{zsum}
\ee
Note that the condition in Eq.~\eqref{eq.NILCfilterreq} guarantees that signal power is preserved in this series of operations.
\end{enumerate}

\subsection{Deprojection of foregrounds: constrained ILC}\label{sec:constrained_ILC}

The weights in Eq.~\eqref{ILC_weights_app} are chosen to minimize the variance in the recovered map. However, sometimes certain foregrounds can significantly bias a signal of interest and a slight increase in variance is an acceptable price to pay for a significant reduction of this bias.  This can be particularly true for cross-correlation measurements using ILC Compton-$y$ maps, as some foregrounds are more highly correlated with the field with which we are cross-correlating than the signal we are trying to isolate.  For example, the CIB is a foreground for the tSZ effect, and is more highly correlated with CMB lensing than the tSZ signal is~\cite{2011MNRAS.410.2481R}.  In such cases, one may choose to build a ``constrained'' ILC map, as described in the following.

If the SED of a foreground signal is known, we can ``deproject'' it from our final map. In particular, let us now say that the signal in the sky is
\be
T_i(\hat n) = a_i s(\hat n) +b_i f(\hat n)+ n^\prime_i(\hat n),
\ee
where $f(\hat n)$ is the foreground we wish to remove and $b_i$ is its SED. If the weights in the linear combination obey the condition
\be
\sum_i b_i w_i = 0,\label{deprojection_condition}
\ee
the resulting signal will not contain $f(\hat n)$. The weights that obey both Eqs.~\eqref{ILC_condition} and~\eqref{deprojection_condition} and result in a minimum-variance estimate of the signal are
\be
w = \mathbf{e}^T\lb \mathbf{A}^T \mathbf{\mathcal{C}}^{-1} \mathbf{A}\rb^{-1} \mathbf{A}^T \mathcal{C}^{-1}
\ee
where the vector $\mathbf{e}^T=[1 \,\,0]$, and the matrix $\mathbf{A} = [a \,\,b]$.  Written explicitly in terms of components, this is
\be
w_i =  \frac{\left(b_k \left(\mathcal C^{-1}\right)_{kl} b_l\right)^{-1} \left(\mathcal {C}^{-1}\right)_{ij} a_j - \left(a_k \left(\mathcal {C}^{-1}\right)_{kl} b_l\right)^{-1} \left(\mathcal {C}^{-1}\right)_{ij} b_j}{\left(a_k \left(\mathcal C^{-1}\right)_{kl} a_l\right)\left(b_m \left(\mathcal C^{-1}\right)_{mn} b_n\right)-\left(a_k \left(\mathcal C^{-1}\right)_{kl} b_l\right)^2} \,.
\ee

\subsubsection*{Multiply constrained ILC}

This can be extended to the case where multiple components are simultaneously deprojected (see, e.g.,~\cite{2021MNRAS.503.2478R,2021PhRvD.103j3510A,2023arXiv230308121K}). Let us consider a case where we deproject $N_{\mathrm{deproj}}$ foregrounds, such that there are $1+N_{\mathrm{deproj}}$ constraints the ILC must obey: the signal-preserving ILC constraint for the signal of interest in Eq.~\eqref{ILC_condition}, along with the nulling condition in Eq.~\eqref{deprojection_condition} for the SEDs of each of the $N_{\mathrm{deproj}}$ components we wish to deproject. Note that there is a maximum number of foregrounds we can deproject, corresponding to $N_{\mathrm{deproj}}=N_{\mathrm{freq}}-1$, as an attempt to add further constraints would over-constrain the system of equations.

In this case, the weights are given by~\cite{2023arXiv230308121K}
\begin{align}
w^i =\left(\mathcal C^{-1}\right) _{ij}\frac{1}{\mathrm{det}\mathcal{Q}}\bigg{(}\mathrm{det}\left(\mathcal Q_{1,2,...,N_{\mathrm{deproj}};1,2,...,N_{\mathrm{deproj}}}\right)\mathcal A_{j 0 } -\mathrm{det}\left(\mathcal Q_{0,2,...,N_{\mathrm{deproj}};1,2,...,N_{\mathrm{deproj}}}\right)\mathcal A_{j 1}+\\\nonumber
+\mathrm{det}\left(\mathcal Q_{0,1,3,...,N_{\mathrm{deproj}};1,2,...,N_{\mathrm{deproj}}}\right)\mathcal A_{j 2}-\mathrm{det}\left(\mathcal Q_{0,1,2,4,...,N_{\mathrm{deproj}};1,2,...,N_{\mathrm{deproj}}}\right)\mathcal A_{j 3}+\cdots \bigg{)},
\end{align}
where $\mathcal Q$ is an $\left(N_{\mathrm{deproj}}+1\right)\times\left(N_{\mathrm{deproj}}+1\right)$-dimensional symmetric matrix with components
\be
\mathcal Q_{\alpha \beta} \equiv \left(\mathcal C^{-1}\right)_{ij}\mathcal A_{i \alpha}\mathcal A_{j \beta},
\ee
and where $\mathcal A_{i \alpha}$ is an  $N_{\mathrm {freq}}\times \left(N_{\mathrm{deproj}}+1\right)$-dimensional matrix with components given by the SEDs of the signals we wish to preserve or deproject, with the $\alpha=0$ column that of the SED we are preserving and the $\alpha=1,...$ components given by the SEDs we are deprojecting (note that the Greek indices $\alpha,\beta,...$ refer to these  $N_{\mathrm{deproj}}+1$ dimensions). The matrix $\mathcal Q_{...; 1,2,...,N_{\mathrm{deproj}}}$ refers to the $N_{\mathrm{deproj}}\times N_{\mathrm{deproj}}$-dimensional sub-matrix of $\mathcal Q$ which is formed by removing the $\gamma$ row and the 0th column, where $\gamma$ is the index that is dropped from the first component list.

\subsection{The ILC bias}\label{sec:ILC_bias}

The ILC bias is a well-known issue in the ILC algorithm~\cite{2009A&A...493..835D}, which is caused by the fact that the covariance matrix is estimated directly from the maps, using only a finite number of modes. Chance fluctuations lead to incorrect estimates of the covariance and can lead to correlations between the weights and the signal of interest. This can be minimized by measuring the covariance over a large enough domain with enough modes that the covariance matrix estimation is immune to chance fluctuations. 

We can define the ILC bias as follows. Recall the signal estimate is
\be
\tilde s = \sum _i w_i T_i = s + \tilde n
\ee
where $\tilde n$ is given by
\be
\tilde n = \sum _i w_i n_i.
\ee
The variance of $\tilde s$ is
\be
\left<\tilde s^2\right> = \left<\left ( s + \tilde n\right)^2 \right> = \left<s^2\right> + 2 \left<s \tilde n\right> + \left<\tilde n^2\right>.
\ee
The contribution from $\left<\tilde n^2\right>$ adds ``noise bias'' to the estimation of a power spectrum from the map, and depends (to first order) on the foregrounds and the noise properties of the maps. By ``ILC bias'', we refer explicitly to the term sourced by the correlation of the signal and the $\tilde n$ term: $\left<s \tilde n\right>$. For weights constructed from the true covariance matrix of the underlying theory (and for cases where the foregrounds and the signal are uncorrelated), this term exactly vanishes. However, even in cases when the foregrounds and the signal are uncorrelated, this term can be non-zero as the weights are constructed from a covariance matrix that is \textit{measured from the data}; thus, the signal appears in $\tilde n$ through $w_i$. We define $b_{\mathrm{ILC}}$ exactly as
\be
b_{\rm ILC} = \left<s \tilde n\right>.
\ee

The fractional size of the ILC bias $\frac{b_{\rm ILC}}{\left<s^2\right>} $  can be estimated from the number of modes used to calculate the covariance matrix, $N_{\mathrm{modes}}$:\footnote{Note that this expression corrects an error in Eq.~3 of Ref.~\cite{2016A&A...594A..22P} --- their numerator should be $(N_{\rm channels} - 2)$, due to the deprojection of the CMB component in their $y$-map.}
\be
\frac{b_{\rm ILC}}{\left<s^2\right>} = \frac{|1+N_{\mathrm{deproj}}-N_{\mathrm{freq}}|}{N_{\mathrm{modes}}} \,.    \label{ILC_bias}
\ee
Note that this bias is negative~\cite{2009A&A...493..835D}, but it will be convenient later in our work to have defined this quantity using an absolute value such that it is positive-definite.  Generically, increasing $N_{\rm modes}$ suppresses this bias.

In a real-space domain $\mathcal {D}^{\mathrm{real}}$, the number of modes $N_{\mathrm{modes}}$ is equal to the number of pixels in $\mathcal {D}^{\mathrm{real}}$. In a harmonic domain $\mathcal {D}^{\mathrm{harm}}$, the number of modes is given by $\sum_\ell (2\ell+1)$, with the sum taken over all $\ell\in \mathcal {D}^{\mathrm{harm}}$ (here, assuming full-sky data, with the number of harmonic modes otherwise reduced by a factor of $f_{\rm sky}$).

In a needlet domain $\mathcal D^{\mathrm{real},I}_{\hat n}$, the number of modes is given by $\frac{N_{\mathrm{pix}}}{N_{\mathrm{pix},\Omega}}\sum_\ell (2\ell+1)\left( h_\ell^I\right)^2$, where $N_{\mathrm{pix}}$ is the (effective) number of real-space pixels in $\mathcal D^{\mathrm{real},I}_{\hat n}$; $N_{\mathrm{pix},\Omega}$ is the number of pixels on the entire sphere; and the sum is taken over all $\ell$, with the needlet filter function $h_\ell^I$ appropriately weighting the contributions from the different multipoles.  Note that this expression is only valid for real-space domains defined by top-hat window functions in pixel space; it is also possible to allow for more complicated real-space domains, such as domains defined by a real-space Gaussian window function, in which case $N_{\mathrm{pix}}$ does not just count pixels but is weighted by the window function defining $\mathcal D^{\mathrm{real},I}_{\hat n}$.

\section{Estimating the tSZ signal}\label{sec:ilc_ymap}

The tSZ spectral distortion is given by
\be
\frac{\Delta T^{\mathrm{tSZ}}(\hat n, \nu)}{T_{\mathrm{CMB}}} = g_\nu y (\hat n),
\ee
where $g_\nu$ is the tSZ spectral function~\cite{1970Ap&SS...7....3S}:
\be
g_\nu =x\coth\lb\frac{x}{2}\rb-4 \,\label{gnu_tsz}
\ee
with $x\equiv\frac{h\nu}{k_B T_{CMB}}$. Here $h$ is Planck's constant, $k_B$ is Boltzmann's constant, $T_{\mathrm{CMB}} = 2.726$ K is the mean temperature of the CMB~\cite{1996ApJ...473..576F,1999ApJ...512..511M,2009ApJ...707..916F}, and $y(\hat n)$ is the dimensionless (and frequency-independent) Compton $y$-parameter that quantifies the integral of the electron pressure along the line of sight (LOS). An ILC map that preserves the SED given by Eq.~\eqref{gnu_tsz} is a map of the Compton-$y$ anisotropies in our Universe.

We construct a set of Compton-$y$ maps from the single-frequency maps of the \textit{Planck} \NPIPE~data release 
(PR4)~\cite{2020A&A...643A..42P}. In our ``standard frequency coverage'' (default) case, we  use the maps at frequencies $\{30,44,70,100,143,217,353,545\}$ GHz, i.e., we use all of the maps from the Low Frequency Instrument (LFI) and all of the maps from the High Frequency Instrument (HFI) except for that at 857 GHz. We also make an extended-frequency-coverage version which includes 857 GHz, and additionally a version which excludes both 545 and 857 GHz. 

The NILC method has been previously applied to \textit{Planck} data to construct $y$-maps, e.g., in the official \textit{Planck} analysis of the PR1 (2013)~\cite{2014A&A...571A..21P} and PR2 (2015) data releases~\cite{2016A&A...594A..22P}, along with other component separation algorithms, in particular the modified internal linear combination (MILCA)~\cite{2016A&A...594A..22P,2022MNRAS.509..300T}. Ref.~\cite{2014JCAP...02..030H} (hereafter HS14) used a harmonic ILC (HILC) to construct a Compton-$y$ map with which to measure the tSZ-CMB lensing cross-correlation, $C_\ell^{y\kappa}$. Notably, Ref.~\cite{2022MNRAS.509..300T} applied MILCA to the \NPIPE\ single-frequency maps to build a lower-noise $y$-map than the 2015 \textit{Planck} map, and similarly Ref.~\cite{2023arXiv230510193C} has recently applied NILC to the \NPIPE\ maps.

In this section we describe our analysis settings for the NILC algorithm.
 In Section~\ref{sec:nilc_filters} we present the harmonic- and real-space filters we use for our NILC. We present the pre-processing steps that we apply before performing the NILC analysis in Section~\ref{sec:preprocessing}.  We discuss our foreground deprojection methods subsequently in Section~\ref{sec:deprojection_choices}.

\subsection{Harmonic- and real-space filter choices for the NILC analysis}\label{sec:nilc_filters}

\subsubsection{Harmonic-space filters}\label{sec:harmonic_filters}

\begin{figure}[t!]
\includegraphics[width=0.49\textwidth]{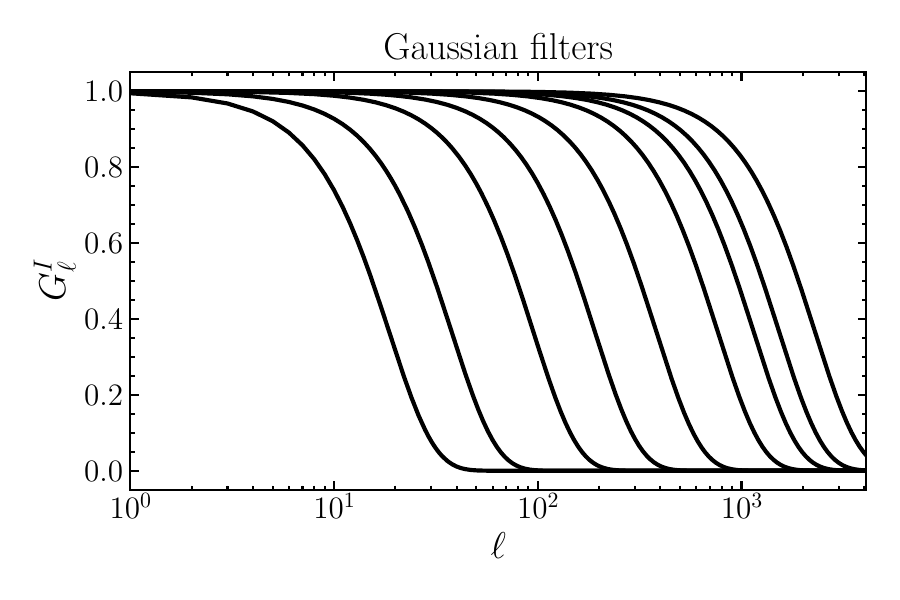}
\includegraphics[width=0.49\textwidth]{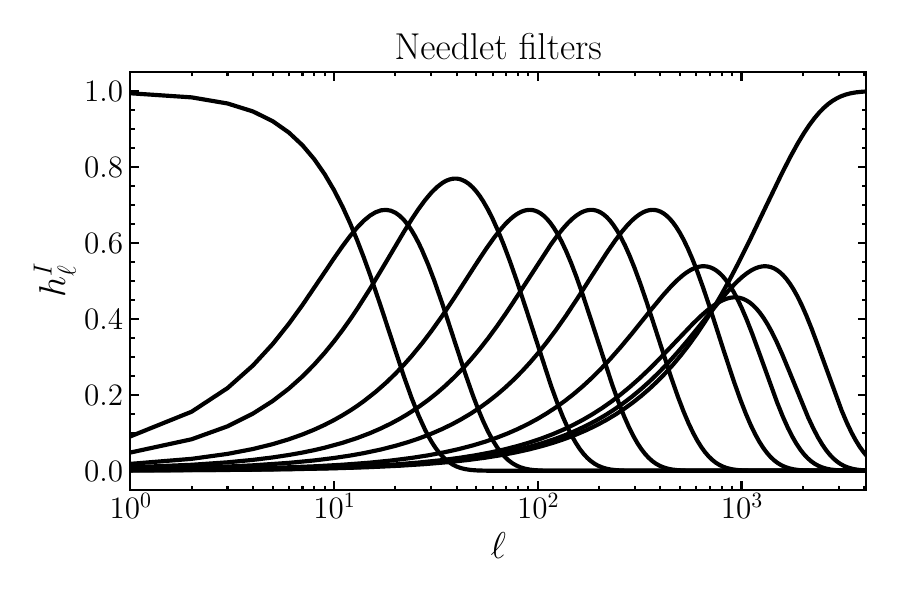}
\caption{The nine Gaussian filter functions (\textit{left}) used to construct the ten needlet harmonic-space filters (\textit{right}) according to Eq.~\eqref{gaussian_to_needlet}.  The FWHMs of the Gaussians are listed in Section~\ref{sec:harmonic_filters} (and repeated in Table~\ref{tab:NILC_details}).}\label{fig:needlet_scales}
\end{figure}

Following the official \textit{Planck} NILC approach to construct Compton-$y$ maps~\cite{2016A&A...594A..22P}, we use $N_{\mathrm{scales}}=10$ Gaussian needlet functions $h^I_\ell$ (where $0<I<N_{\mathrm{scales}}$) with scales corresponding to full-width-at-half-maximum (FWHM) values of $\{600, 300, 120, 60, 30, 15, 10, 7.5, 5\}$ arcmin.  To construct these needlet filters, we first define Gaussian filters $G^I_\ell$ for each scale $0<I<N_{\mathrm{scales}}$, then construct the needlet filters according to
\begin{align}
h^I_\ell =\begin{cases}
 G^I_\ell & I=0;\\
 \sqrt{\left(G^I_\ell{}\right)^2-\left(G^{I-1}_\ell{}\right)^2}& 0<I<N_{\mathrm{scales}}-1;\\
 \sqrt{1-\left(G_{\ell}^I\right)^2}& I=N_{\mathrm{scales}}-1.\label{gaussian_to_needlet}
\end{cases}
\end{align}
The Gaussian filters are defined as usual according to 
\be
G_\ell^I = \exp\lb-\frac{\ell\lb\ell+1\rb}{16\ln 2}\left(\Theta_{\mathrm{FWHM}}^I\right)^2\rb,
\ee
where $\Theta_{\mathrm{FWHM}}^I$ is the beam FWHM appropriate for scale $I$ in radians.   We require the needlet filters to obey the unit-transmission criterion~\cite{2009A&A...493..835D}
\be
\sum _I \left(h^I_\ell\right)^2 = 1 \quad \forall \, \ell \,.
\ee
The Gaussian filters $G^I_\ell$ and the needlet filters $h^I_\ell$ are shown in Figure~\ref{fig:needlet_scales}.

Before performing the needlet decomposition of the single-frequency maps, we convolve them to a common beam. However, note that, while we perform the NILC with many input frequency maps, the lowest-frequency maps have much larger beam FWHMs than the highest-frequency maps, and the signal is exponentially sub-dominant to the instrumental noise on small-to-intermediate scales (after beam deconvolution). Thus, in practice, we drop particular frequency maps from the NILC on the needlet scales where we expect no information from these maps. In particular, we define a ``beam threshold criterion'' $b_{\rm thresh}$, which we take to be $b_{\rm thresh} = 10^{-3}$. If, for a given needlet scale, the beam of a given frequency channel reaches this threshold at a lower $\ell$ than the needlet filter does for that scale, we drop that frequency channel from the NILC.  Mathematically: we find the multipole $\ell_h$ where $h^I_{\ell_h} = b_{\rm thresh}$ (on the decreasing side of the needlet filter) and the multipole $\ell_b$ where $b^i_{\ell_b} = b_{\rm thresh}$.  We then require that $\ell_h < \ell_b$ in order for a frequency map to be included in the NILC at this needlet scale. If this criterion is not satisfied for the beam of a given frequency channel, then that channel is dropped from the NILC at this needlet scale.  Note that this procedure assumes monotonicity of the beam function, and also assumes that the filter function has a peak (and thus a decreasing region).  The latter assumption does not hold for the smallest-scale needlet filter (see Figure~\ref{fig:needlet_scales}), so for this scale we simply use the same maps as used at the penultimate needlet scale.

\begin{table}
\begin{tabular}{|c |c |}\hline
Needlet scale & Frequencies included\\\hline\hline
0 &$\left\{30,44,70,100,143,217,353,545,(857)\right\}$ GHz\\\hline
1 &$\left\{30,44,70,100,143,217,353,545,(857)\right\}$ GHz\\\hline
2 &$\left\{30,44,70,100,143,217,353,545,(857)\right\}$ GHz\\\hline
3 &$\left\{30,44,70,100,143,217,353,545,(857)\right\}$ GHz\\\hline
4 &$\left\{44,70,100,143,217,353,545,(857)\right\}$ GHz\\\hline
5 &$\left\{70,100,143,217,353,545,(857)\right\}$ GHz\\\hline
6 &$\left\{100,143,217,353,545,(857)\right\}$ GHz\\\hline
7 &$\left\{143,217,353,545,(857)\right\}$ GHz\\\hline
8 &$\left\{143,217,353,545,(857)\right\}$ GHz\\\hline
9 &$\left\{143,217,353,545,(857)\right\}$ GHz\\\hline

\end{tabular}
\caption{The frequency maps included in our NILC algorithm at each needlet scale, given our beam threshold criterion of $10^{-3}$ (see Section~\ref{sec:harmonic_filters}).  The 857 GHz channel is listed in parentheses as it is not included in our default NILC pipeline, but only in some variations thereof.}
\label{tab:beam_criterion_Scales}
\end{table}

\subsubsection{Real-space filters}

Our real-space domains $\mathcal D^{\mathrm{real},I}_{\hat n}$ (on which the frequency-frequency covariances are computed) are also defined by Gaussian kernels. We define their FWHM values by calculating the number of modes $N_{\mathrm{modes}}$ required to ensure that the fractional ILC bias (defined in Eq.~\eqref{ILC_bias}) is always below a tolerance $b^{\mathrm{tol}}$, which we set to $b^{\mathrm{tol}} = 0.01$:
\be
\label{eq.ILCbias_criterion}
\frac{b_{\rm ILC}}{\left<s^2\right>} < b^{\mathrm{tol}} \,,
\ee
where $\left<s^2\right>$ is the variance of the signal on the domain $\mathcal D^{\mathrm{real},I}_{\hat n}$. At each needlet scale $I$, we convert the value of $N_{\mathrm{modes}}$ implied by Eqs.~\eqref{ILC_bias} and~\eqref{eq.ILCbias_criterion} into FWHMs for the real-space Gaussian filters by first calculating the total number of modes on the full sky at each needlet scale $I$ according to 
\be
N_{\mathrm{modes}}^I = \sum_\ell \left( 2\ell+1\right) \left(h^I_\ell\right)^2.
\ee
Noting that the fractional sky area covered by the real-space Gaussian is $2\pi \sigma_{\mathrm{real},I}^2/(4\pi) = \sigma_{\mathrm{real},I}^2/2$, we then define a variance $\sigma_{\mathrm{real},I}^2$ for our real-space filters via 
\be
\sigma_{\mathrm{real},I}^2 = 2  \left( \frac{|1+N_{\mathrm{deproj}}-N_{\mathrm{freq}}|}{b^{\mathrm{tol}}N^I_{\mathrm{modes}}} \right)
\ee
and convert this into a FWHM with
\be
\theta_{\mathrm{FWHM},I}^2 = 8\ln 2 \, \sigma_{\mathrm{real},I}^2 \,.
\ee
The resulting real-space FWHM values that we use to construct our NILC maps are listed in Table~\ref{tab:NILC_details}; note that they in principle depend on the number of foregrounds deprojected $N_{\mathrm{deproj}}$, with smaller real-space FWHMs allowed when deprojecting more components (at fixed $b^{\rm tol}$). However, in practice, we compute the covariance matrices only \textit{once}, using the filters of the $N_{\mathrm{deproj}}=0$ case, as the covariance matrix calculation and inversion is the most computationally intensive part of the NILC.  We then save the covariance matrices and inverse covariance matrices for use in the remaining NILC calculations, for different deprojection choices. As the real-space domains used in the $N_{\mathrm{deproj}}=0$ case are larger than those required to satisfy the ILC bias threshold for $N_{\mathrm{deproj}}>0$, the ILC bias threshold will be satisfied in all cases. Note that, in practice, $N_{\mathrm{freq}}$ depends on the needlet scale, as we drop the lower-resolution maps from the higher-resolution needlet scales according to Table~\ref{tab:beam_criterion_Scales}.

\begin{table}[t!]
\begin{tabular}{|c|c||c|c|c|c|c|c|c|c|c|c|}\hline
\multicolumn{2}{|c||}{Needlet scale number $I$}& 0 & 1 & 2 &  3&  4&  5&  6&  7& 8 & 9\\\hline
\multicolumn{2}{|c||}{Needlet scale FWHM (arcminute) }& 600 & 600-300 & 300-120 &  120-60&  60-30&  30-15&  15-10&  10-7.5&  7.5-5 & 5\\\hline\hline
\multirow{3}{*}{Real-space FWHMs (degrees)}&$N_{\mathrm{deproj}}=0$  & 373.8  & 216.0 &  81.65&   43.20 &  20.00& 
   9.13 &   6.32 &   4.63 &   2.74   & 0.89\\\cline{2-12}
&$N_{\mathrm{deproj}}=1$  & 346.1&  200.0 &  75.59  & 40.00 &  18.26& 
   8.16 &   5.48  &  3.78&    2.24  &  0.72\\\cline{2-12}
&$N_{\mathrm{deproj}}=2$  & 315.9 & 182.6&   69.01 &  36.51 &  16.33& 
   7.07 &   4.47 &   2.67 &   1.58  &  0.51\\\hline
\end{tabular}
\caption{Details of our NILC pipeline, including the FWHMs of the Gaussians used to construct the harmonic needlet filters as described in Section~\ref{sec:harmonic_filters}, and the FWHMs of the real-space filters used to define the domains on which we calculate the real-space covariance matrices. The real-space domain size depends on the number of deprojected foregrounds $N_\mathrm{deproj}$, as indicated. Note that we do not actually use the real-space filters specified here for $N_{\mathrm{deproj}}>0$, as the larger filters calculated for $N_{\mathrm{deproj}}=0$ will automatically satisfy the ILC bias threshold in these cases.  }\label{tab:NILC_details}
\end{table}

Once we have defined the domains $\mathcal {D}_{\hat {n} }^{\mathrm{real},I}$, we calculate the mean and the frequency-frequency covariance of the needlet coefficients $T_i^I(\hat n)$ on them by smoothing each quantity with a Gaussian beam of the appropriate FWHM.  In particular, the mean is simply given by the smoothed needlet coefficient maps, while the covariance is calculated by subtracting these means from the full needlet coefficient maps and multiplying them together, then smoothing the result.

 \subsection{Pre-processing details}\label{sec:preprocessing}

Before applying the NILC algorithm to the data, we apply a mask to each single-frequency map to remove very bright Galactic emission and point sources. This prevents the dominant emission from these regions (which, even if retained, would be masked in the final analysis) from significantly affecting the ILC weights calculated slightly away from (but in the vicinity of) these regions, and thus allows for better component separation away from these regions.

\begin{figure}[t!]
\includegraphics[width=0.32\textwidth]{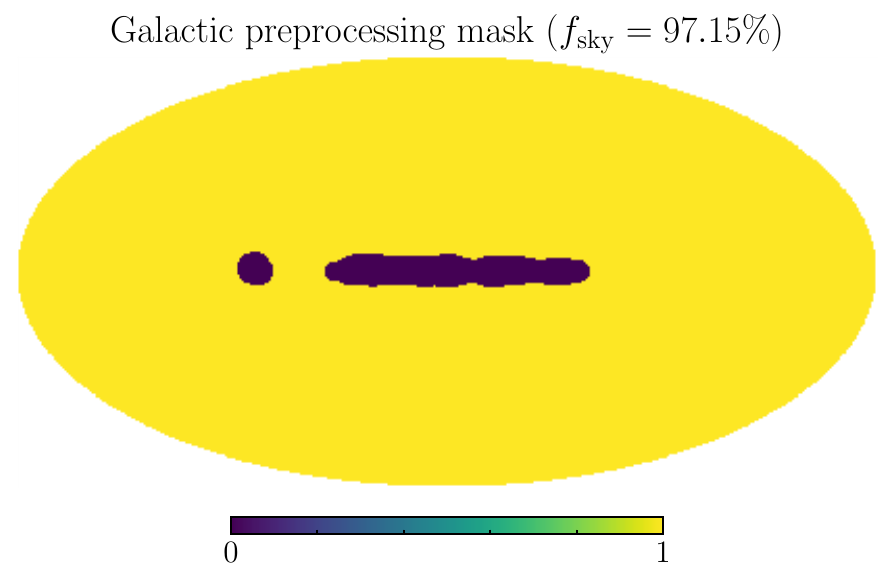}
\includegraphics[width=0.32\textwidth]{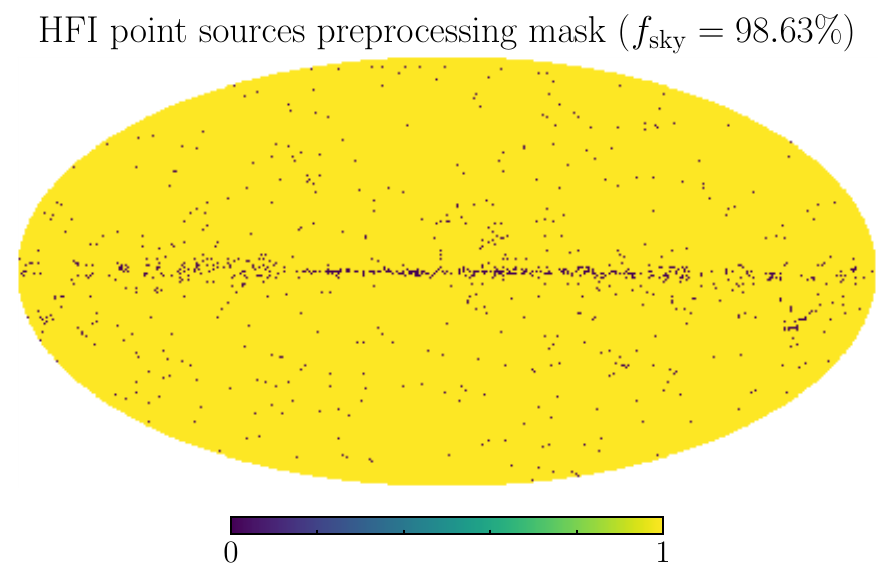}
\includegraphics[width=0.32\textwidth]{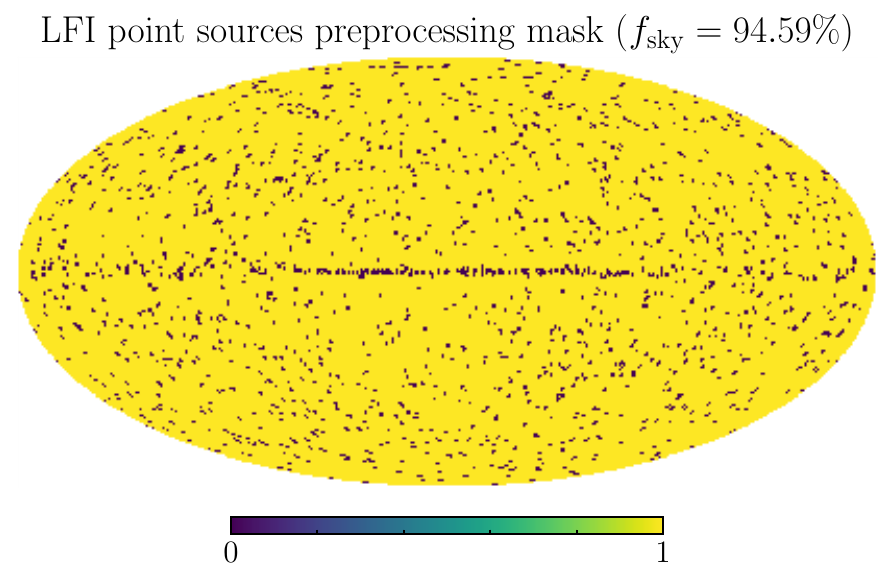}

\caption{The masks used in the preprocessing of our needlet ILC pipeline. These masks are applied to the single-frequency maps which are then inpainted by iteratively replacing each masked pixel with the mean of the surrounding unmasked pixels (neighbors for an unmasked pixel) before we run the needlet ILC algorithm.}
\label{fig:preprocessing_masks}
\end{figure}

We inpaint the regions of each map covered by these masks using a diffusive inpainting scheme, which iteratively replaces masked pixels with $\ge4$ unmasked neighbors by the mean of the surrounding unmasked pixels. The Galactic mask we use is the same as that used in the pre-processing of the \textit{Planck} 2015 \texttt{NILC} tSZ map~\cite{2016A&A...594A..22P}, which we download from the \href{https://pla.esac.esa.int/#home}{Planck Legacy Archive} (PLA)\footnote{This is available at {\url{http://pla.esac.esa.int/pla/aio/product-action?MAP.MAP_ID=COM_CompMap_Compton-SZMap-nilc-ymaps_2048_R2.00.fits}} (field $=3$).}. The point source mask we use is the point source catalogue mask from \textit{Planck}, which we also download from the PLA, using the low-frequency catalogue for the LFI maps and the high-frequency catalogue for the HFI maps\footnote{These are available at \url{http://pla.esac.esa.int/pla/aio/product-action?MAP.MAP_ID=HFI_Mask_PointSrc_2048_R2.00.fits
} (HFI) and \url{http://pla.esac.esa.int/pla/aio/product-action?MAP.MAP_ID=LFI_Mask_PointSrc_2048_R2.00.fits
} (LFI).}. These preprocessing masks are shown in Figure~\ref{fig:preprocessing_masks}. The Galactic mask covers 2.85\% of the sky; the HFI point source mask covers 1.37\% of the sky; and the LFI point source mask covers 5.41\% of the sky. The combination of the Galactic mask and the HFI point source mask covers 3.93\% of the sky and the combination of the Galactic mask and the LFI point source mask covers 7.94\% of the sky; the combination of all three covers 8.50\% of the sky, which at the end defines the total masked sky area of our final NILC maps.

As the \NPIPE~single-frequency maps retain the dominant contribution from the kinematic dipole, before inpainting we remove the kinematic dipole as measured by the Commander component separation algorithm~\cite{2020A&A...643A..42P}, and then subtract the remaining monopoles of the maps. Finally, following Ref.~\cite{2016A&A...594A..22P}, we deconvolve the beams of each map (the beams of the \textit{Planck} maps are listed in Ref.~\cite{2020A&A...641A...1P}; we repeat them in Table~\ref{tab:fwhm}) and reconvolve all maps to a common beam of $10^\prime$. Note that, as the covariances are calculated in real space, the beam at which the NILC is performed can significantly affect the final result, as the needlet filters are so broad in scale that the large variance of the small-scale modes can contribute significantly if they are not sufficiently beam-convolved. Thus the weights can spuriously adapt to mitigate the variance sourced by the smallest-scale modes in these frequency channels. Such an impact could be mitigated by choosing needlet scales with less broad coverage in $\ell$-space.  In this work, however, we choose to follow the official \textit{Planck} analysis and perform the NILC on maps at $10^\prime$ resolution.

\section{Deprojection of various components}\label{sec:deprojection_choices}

In Section~\ref{sec:constrained_ILC}, we discussed constrained ILC and foreground deprojection. In this section we present our specific choices for foreground deprojection, in particular for removing the CMB and the CIB.

\subsection{Deprojection of the CIB}
\label{subsec:CIBSED}

For many cross-correlation analyses of the tSZ signal with large-scale structure (LSS) tracers, it is necessary to deproject the CIB, which is also correlated with LSS and which can bias a $\langle y \times LSS \rangle$ measurement if unmitigated. 
However, unlike the tSZ and CMB signals, which display no frequency decorrelation and which have SEDs that are well understood from first principles and can thus be calculated theoretically, the CIB is not described perfectly by one SED. It is sourced by the line-of-sight integrated thermal emission of different objects; in particular, different frequency channels are sensitive to slightly different objects, as source emission at different redshifts will be redshifted into different frequency bands. This leads to frequency decorrelation between the CIB channels.  However, as the correlation coefficients are $\gtrsim 90$\% at the frequencies of interest~\cite{2013ApJ...772...77V,2014A&A...571A..30P,2017MNRAS.466..286M,2019ApJ...883...75L}, it is still possible to clean the CIB using multi-frequency measurements. Its SED does not need to be known for the unconstrained ILC; however, if we wish to explicitly deproject it, we must model its SED. 

Let us explicitly write the CIB intensity as
\be
I_\nu^{\mathrm{CIB}}( \hat n) = \Theta_\nu ^{\mathrm{CIB}}A^{\mathrm{CIB}} (\hat n) \,,
\label{CIB_intensity2}
\ee
where all frequency dependence is absorbed in the CIB SED, ${\Theta_\nu^{\mathrm{CIB}}}$, and $A^{\mathrm{CIB}}(\hat n)$ is a template of the CIB anisotropies.  Here we model the CIB as a modified blackbody, with an SED given by \be
 \Theta_\nu^{\mathrm{CIB}}\propto\nu^\beta B_\nu(T^{\mathrm{eff}}_{\mathrm{CIB}})\label{CIB_SED}
\ee
where $B_\nu(T)$ is the Planck function
\be
B_\nu(T) = \frac{2 h \nu^3}{c^2}\frac{1}{e^{\frac{h \nu}{k_BT}}-1} \,. \label{mbb}
\ee
The SED depends on two parameters: the spectral index $\beta$ and the effective CIB temperature $T^{\mathrm{eff}}_{\mathrm{CIB}}$. We stress that $T^{\mathrm{eff}}_{\mathrm{CIB}}$ is not a physical temperature, but a parameter in this effective description.  Also note that the overall normalization of the CIB SED in Eq.~\eqref{CIB_SED} is arbitrary\footnote{One only needs to specify the overall normalization for the component that is preserved in an ILC map, so that the output map is in the correct units; the components that are deprojected do not need their normalization specified.}. For concreteness, we will thus write the CIB intensity as
\be
I_\nu^{\mathrm {CIB}}(\hat n) = \Theta_\nu ^{\mathrm{CIB}}A^{\mathrm {CIB}}(\hat n) = \left(\frac{\nu}{\nu_0}\right)^{\beta+3}\frac{1}{e^{x_{\mathrm{CIB}}}-1} A^{\mathrm {CIB}}(\hat n)\label{CIB_intensity}
\ee
where $A^{\mathrm {CIB}}(\hat n)$ is the underlying CIB component template, and with $x_{\mathrm{CIB}}\equiv\frac{h \nu}{k_B T^{\mathrm{eff}}_{\mathrm{CIB}}}$.  We introduce a pivot frequency $\nu_0$ in Eq.~\eqref{CIB_intensity}, {which one can choose arbitrarily; we set $\nu_0 = 353$ GHz.} It is important to note that this quantifies the CIB intensity, not temperature; natural units for the CIB intensity are Jy/sr or MJy/sr. As we analyze our maps in units of CMB thermodynamic temperature $\mu \mathrm{K}$, we must convert from intensity to  $\mu \mathrm{K}$ in order to write the SED of the CIB as is relevant for our maps. This is a frequency-dependent conversion, which is obtained by differentiating the Planck function (Eq.~\eqref{mbb}) with respect to $T$ and is given by
\be
\left(\frac{dB_\nu}{dT}\right)_{T=T_{\rm CMB}} = \frac{2 h \nu^3}{c^2}\frac{e^x}{\left(e^x-1\right)^2}\frac{x}{T_{\mathrm{CMB}}},
\ee
where $x\equiv \frac{h\nu}{k_B T_{\mathrm{CMB}}}$.

The parameters of the effective CIB SED ($\beta,T^{\mathrm{eff}}_{\mathrm{CIB}}$) can be fit to observations of the CIB. For use in our fiducial CIB deprojections, we use a determination of the CIB SED determined by a fit to the monopole predictions of the best-fit halo model of Ref.~\cite{2014A&A...571A..30P}, in particular their Table 10. We reproduce these predictions for the CIB SED in Table~\ref{tab:CIB_monopole}. 

\begin{table}[t!]
\begin{tabular}{|c||c|c|c|c|}\hline
Frequency [GHz]& 217 & 353 & 545 & 857 \\\hline\hline
$\nu I_\nu [\mathrm{n W}/\mathrm{m}^2/\mathrm{sr}]$&$0.077\pm0.003$&$0.53\pm0.02$&$2.3\pm0.1$&$7.7\pm0.2$\\\hline
\end{tabular}
\caption{The predicted values of the CIB monopole in Ref.~\cite{2014A&A...571A..30P}, as calculated from the halo model fit to the CIB power spectra in that work at $\nu=\{217,353,545,857\}$ GHz.}\label{tab:CIB_monopole}
\end{table}

We write a simple Gaussian likelihood for the data in Table~\ref{tab:CIB_monopole} with $1\sigma$ errors given by the quoted error bars, and no covariance between the different frequencies:
\begin{equation}
-2 \ln \mathcal L (\beta, T^{\mathrm{eff}}_{\mathrm{CIB}}, A) = \sum_\nu  \frac{\left(A\Theta_\nu^{\mathrm{CIB}}(\beta, T^{\mathrm{eff}}_{\mathrm{CIB}})- d_\nu\right)^2}{\sigma_\nu^2}\label{likelihood_betaT}
\end{equation}
where $d_\nu$ are the data points in Table~\ref{tab:CIB_monopole} (with the units appropriately converted), $\sigma_\nu$ are the $1\sigma$ errors reported in Table~\ref{tab:CIB_monopole}, and $\Theta_\nu^{\mathrm{CIB}}(\beta, T^{\mathrm{eff}}_{\mathrm{CIB}})$ is the quantity defined in Eq.~\eqref{CIB_SED}.  Note that the expression $A \Theta_\nu^{\mathrm{CIB}}$ is the monopole analogue of Eq.~\eqref{CIB_intensity2}. 

\begin{figure}[t!]
\includegraphics[width=0.5\textwidth]{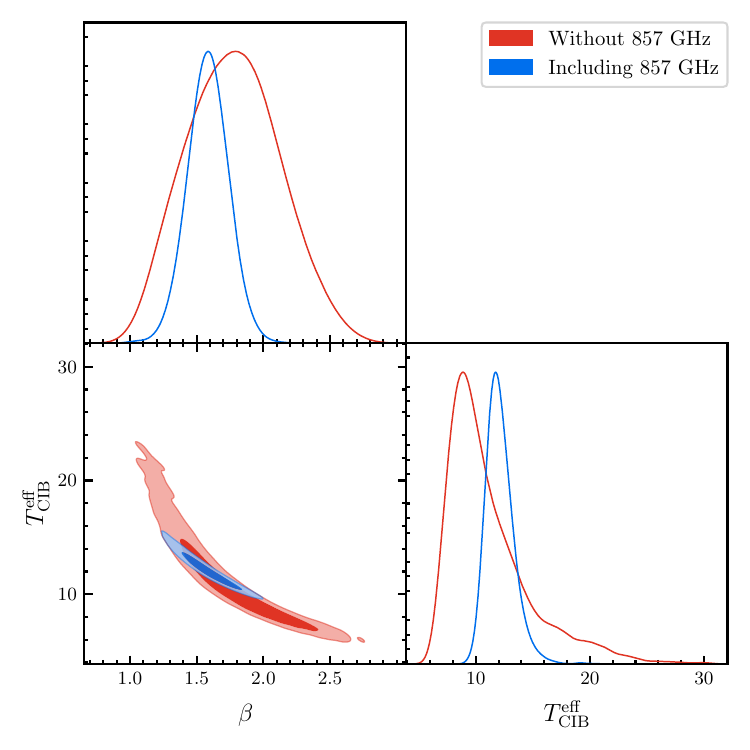}
\caption{The posterior for the parameters of the inferred effective CIB SED. There is significant degeneracy between the parameters $\beta$ and $T_{\mathrm{CIB}}^{\mathrm{eff}}$. 
For the without-857-GHz case, the likelihood is maximized at $\beta=\bestfitbetastandard,\,T_{\mathrm{CIB}}^{\mathrm{eff}}=\bestfitTstandard\,\mathrm{K}$.}\label{fig:posterior_betaT}
\end{figure}

The parameters that maximize the likelihood are found by straightforwardly minimizing {the right-hand side of } Eq.~\eqref{likelihood_betaT}. When we only include information from the three lower frequencies $\nu=\{217,353,545\} \,\mathrm{GHz}$ in the likelihood, the maximum-likelihood parameters are given by $\beta=\bestfitbetastandard,\,T_{\mathrm{CIB}}^{\mathrm{eff}}=\bestfitTstandard\,\mathrm{K}$; when we additionally include 857 GHz, the best-fit parameters are $\beta=\bestfitbetaplus,\,T_{\mathrm{CIB}}^{\mathrm{eff}}=\bestfitTplus\,\mathrm{K}$. To quantify the uncertainty, we perform Markov-chain Monte Carlo (MCMC) sampling of the posterior, using flat, linear priors on all parameters, resulting in a posterior region shown in Figure~\ref{fig:posterior_betaT}. For this, we use the Metropolis-Hastings algorithm implementation of~\cite{2002PhRvD..66j3511L,2013PhRvD..87j3529L} implemented in \texttt{Cobaya}\footnote{\url{https://cobaya.readthedocs.io/en/latest/}}~\cite{2019ascl.soft10019T,Torrado:2020dgo}. We run the chains until they are converged with a Gelman-Rubin criterion~\cite{1992StaSc...7..457G} of $|R-1| < 0.005$. The marginalized means and posteriors of $\beta$ and $T^{\mathrm{eff}}_{\mathrm{CIB}}$ are shown in Table~\ref{tab:betatbestfittable}, along with the best-fit values. 

We stress that these are not intended to be highly accurate posteriors for the CIB SED parameters, as we have not quantified appropriately the covariance of the data points. 
Instead, they are intended to give an idea of a realistic region of variation of the parameters. Note that for the case when we do not include 857 GHz, we are fitting three data points with three parameters, and so our reduced $\chi^2$ is ill-defined and we do not have a proper notion of goodness-of-fit (and, as expected, the $\chi^2$ approaches 0). For the case when we do include 857 GHz, we find that the model is a good fit, as quantified by the reduced $\chi^2$.

\begin{table}
\begin{tabular}{|c||c|c||c|c|}\hline
&\multicolumn{2}{c||}{Without 857 GHz}&\multicolumn{2}{c|}{With 857 GHz}\\\hline\hline
&Mean&Best fit & Mean & Best fit\\\hline
$\beta$&$1.75^{+0.39}_{-0.37}$&$\bestfitbetastandard$&$1.60\pm0.16$&$\bestfitbetaplus$\\\hline
$T^{\mathrm{eff}}_{\mathrm{CIB}}$&$11.87^{+0.75}_{-4.94}$ K & $\bestfitTstandard$ K & $12.01^{+0.99}_{-1.38}$ K & $\bestfitTplus$ K\\\hline
$\chi^2/{\mathrm{dof}}$ & & $1.53\times10^{-15}$/(3-3) && 0.302 / (4-3)\\\hline
\end{tabular}
\caption{The marginalized means and best-fit values of the $\beta$, $T^{\mathrm{eff}}_{\mathrm{CIB}}$ parameters. We also indicate the reduced $\chi^2$ by explicitly writing the calculated $\chi^2$ and the number of degrees of freedom, which is equal to the number of data points minus the number of fitted parameters (recall that we fit an amplitude along with $\beta$ and $T_{\mathrm{eff}}^{\mathrm{CIB}}$).}\label{tab:betatbestfittable}
\end{table}

\begin{figure}[t!]
\includegraphics[width=0.49\textwidth]{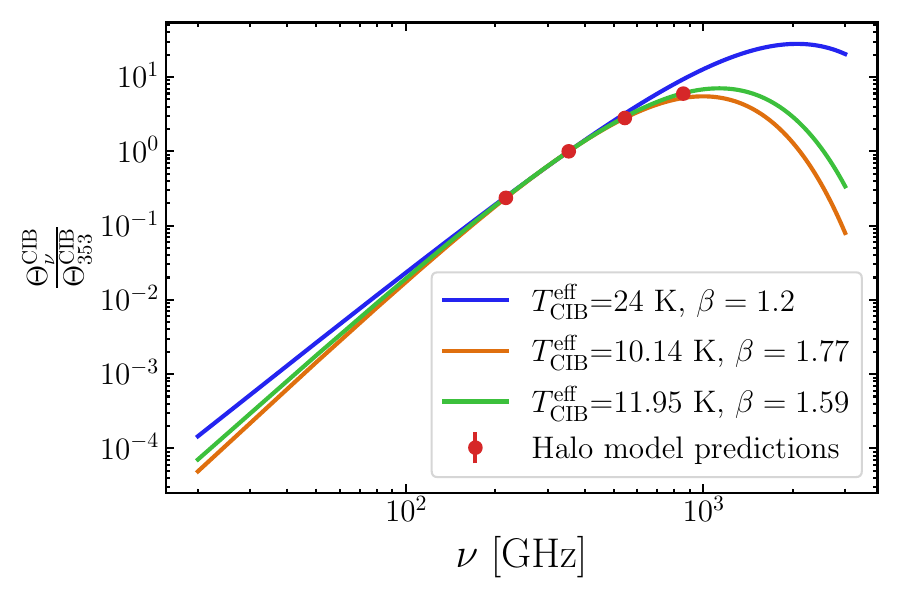}
\includegraphics[width=0.49\textwidth]{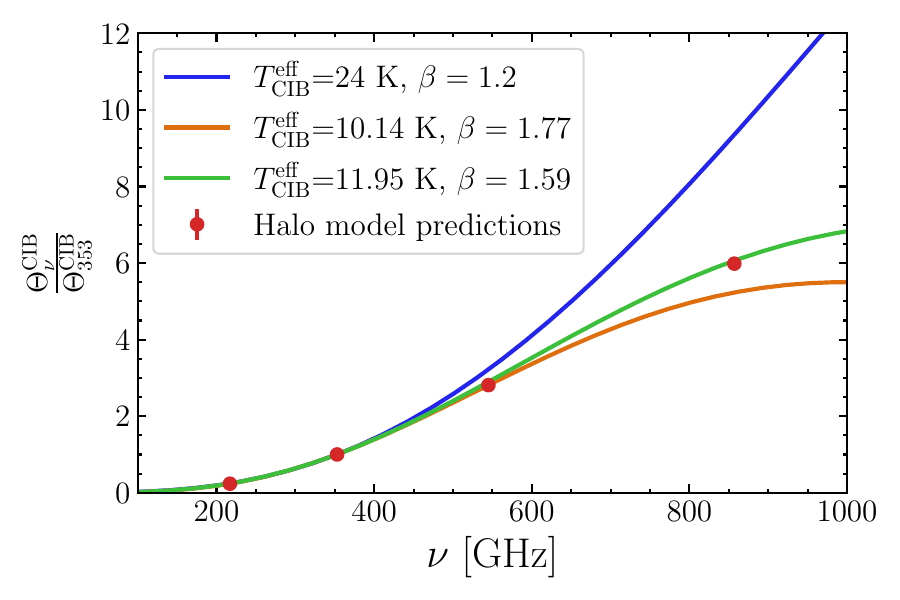}
\caption{The best-fit CIB SEDs, both including the 857 GHz data (green line, $(T_{\mathrm{CIB}}^{\mathrm{eff}},
\beta)=(\bestfitTplus\, \mathrm{K},\bestfitbetaplus)$) in the fit and excluding it (orange line, $(T_{\mathrm{CIB}}^{\mathrm{eff}},
\beta)=(\bestfitTstandard\, \mathrm{K},\bestfitbetastandard)$). We also include for comparison the SED used in Ref.~\cite{2020PhRvD.102b3534M} ($(T_{\mathrm{CIB}}^{\mathrm{eff}},
\beta)=(24\, \mathrm{K},1.2)$), which slightly over-predicts the 545 GHz emission and severely over-predicts the 857 GHz emission, but which is compatible with the data at the $1-2\sigma$ level (see the posteriors in Figure~\ref{fig:posterior_betaT}).  The left and right panels show the same data and theory curves, but with a logarithmic (linear) y-axis on the left (right). }
\label{fig:CIB_SED_bestfit}
\end{figure}

The best-fit SEDs are plotted in Figure~\ref{fig:CIB_SED_bestfit}, along with the SED previously used in  Ref.~\cite{2020PhRvD.102b3534M} to deproject the CIB, which used $T_{\mathrm{CIB}}^{\mathrm{eff}} = 24\,\mathrm{K}$ and $\beta=1.2$.  The latter SED slightly over-predicts the 545 GHz emission and severely over-predicts the 857 GHz emission, although, note that the highest frequency channel used in Ref.~\cite{2020PhRvD.102b3534M} was the \emph{Planck} 545 GHz channel (857 GHz was not used). 
Finally, note that using these (approximate) CIB SEDs requires extrapolation to lower frequencies ($< 217$ GHz) where the SED is not directly constrained.

\subsubsection*{{A note on the best-fit $\beta$, $T_{\mathrm{CIB}}^{\mathrm{eff}}$ and the halo model parameters}}

{Often, the CIB emission is modelled with a halo model with certain parameters that can be fit to data. Many of these halo models model the SED of objects at a given redshift as a modified blackbody, and include in the model a functional form for the (possibly $z$-dependent) physical dust temperature $T_D$ and spectral index $\beta_D$ (see, e.g.,~\cite{2012MNRAS.421.2832S,2021PhRvD.103j3515M}). Often $T_D(z)$ is parameterized as follows:}
\begin{equation}
T_D(z) = T_0(1+z)^\alpha,
\end{equation}
{with $T_0$ and $\alpha$ free parameters that are fitted in the analysis. For example, the halo model analysis of~\cite{2014A&A...571A..30P} found $T_0 = 24.4\pm1.9 \,\,\mathrm{K}$ and $\alpha=0.36\pm0.05$, along with $\beta_D=1.75\pm0.06$.}

{If one wants to directly use these parameters to deproject the CIB sourced at a given redshift, it is important to account for the appropriate redshifting of the CIB temperature. For example, to deproject the CIB emission from $z=1$, one can calculate the temperature of the modified blackbody at $z=1$ as $31.31 \,\,\mathrm{K}$, and then redshift it to today: $T(z)/(1+z)=15.66 \,\,\mathrm{K}$ ($\beta_D$ is unaffected by the redshifting operation and can be directly used in the CIB SED). For $\alpha=1$ (as for the evolution of the CMB temperature), this would be equivalent to always using $T_0$.  The CIB monopole is sourced at $z \approx 2$, which would predict a redshifted temperature of $\approx 12.1 \,\,\mathrm{K}$, not far from our best-fit effective temperatures.
}

\subsection{Deprojection of the first moments of the CIB}

There is significant uncertainty on the parameters ($\beta$, $T_{\mathrm{CIB}}^{\mathrm{eff}}$) of the CIB SED; additionally, it is not a perfect approximation to the CIB emission to describe it as an exact modified blackbody described by a single set of these parameters. For cross-correlations of the tSZ effect with tracers that are more highly correlated with the CIB than with the tSZ field itself (e.g., the CMB lensing potential), imperfect removal of the CIB due to a deprojection with the ``incorrect'' SED can result in a CIB residual that is still a significant bias to the measurement. This systematic must therefore be considered and mitigated when measuring such a cross-correlation, such as in our companion paper~\cite{Paper2} where we measure the tSZ -- CMB lensing cross-correlation.

In Ref.~\cite{2017MNRAS.472.1195C}, a moment-based approach for the description and removal of mm-wave foregrounds was introduced. In particular, their method was designed to mitigate effects due to SED uncertainty or variation of SED parameters due to either spatial averaging within an instrument's beam or along the line-of-sight; the latter is particularly important for foregrounds such as the CIB, which is not emitted at only one redshift but in fact has a wide redshift kernel and is a superposition of imperfect modified blackbodies at various redshifts.

In this approach, the underlying (fundamental) SED is parametrized by some free parameters $\mathbf{p}$, such as a modified blackbody parametrized by $(\beta,T)$.  One then considers variations of these parameters within the instrument beam (and/or sky map pixel) or along the line-of-sight, both of which generically lead to deviations of the ``measured'' SED away from the fundamental SED form.  The observed SED can be represented as a Taylor expansion (in the parameters $\mathbf{p}$) around a point in parameter space $\bar {\mathbf{p}}$, and more accurately modeled using this Taylor expansion.  The zeroth-order moment of the observed SED is the average SED, which can be written as the fundamental SED evaluated at  $\bar {\mathbf{p}}$.  If there were no variations in the SED parameters, then no further modeling would be needed.  However, in general such variations lead to higher-order terms (moments) in the Taylor expansion. Then, the more appropriate SED to use for deprojection in a constrained ILC would be not only the SED evaluated at  $\bar {\mathbf{p}}$, but the entire Taylor expansion. In particular, the SED and its first (and higher-order) derivatives with respect to $ {\mathbf{p}}$ (evaluated at $\bar {\mathbf{p}}$) can be considered as independent components to be deprojected in a constrained ILC. Of course, given a \textit{finite} number of frequency channels, we do not have the freedom to deproject an infinite number of components, but with the multi-frequency coverage of \textit{Planck} it is possible to deproject several components.

We will use this approach to deproject the CIB in a robust manner from our $y$-map.  In particular, we deproject the first moment of the CIB with respect to $\beta$, which in practice amounts to deprojecting an additional component with an SED given by $\frac{\partial I_\nu^{\mathrm{CIB}}}{\partial \beta}$, and also the first moment of the CIB with respect to $T^{\mathrm{eff}}_{\mathrm{CIB}}$, which similarly amounts to deprojecting an additional component with an SED given by $\frac{\partial I_\nu^{\mathrm{CIB}}}{\partial T^{\mathrm{eff}}_{\mathrm{CIB}}}$.

These moments are given explicitly by
\begin{align}
\frac{\partial I_\nu^{\mathrm{CIB}}(\hat n)}{\partial \beta}  =& \ln \left(\frac{\nu}{\nu_0}\right)\, I_\nu^{\mathrm{CIB}}(\hat n) \,; \label{dbeta_moment}\\
\frac{\partial I_\nu^{\mathrm{CIB}}(\hat n)}{\partial T^{\mathrm{eff}}_{\mathrm{CIB}}} =& I_\nu^{\mathrm{CIB}}(\hat n)\frac{x_{\mathrm{CIB}}}{T^{\mathrm{eff}}_{\mathrm{CIB}}}\frac{ e^{x_{\mathrm{CIB}}}}{e^{x_{\mathrm{CIB}}}-1} \,.
\end{align}
For concision, we sometimes refer to these moment components as $\delta\beta$ and $\delta T_{\mathrm{CIB}}^{\mathrm{eff}}$ respectively.
Note that, unlike the SED of the CIB itself, the moment expansion of $\beta$ depends explicitly on the chosen pivot frequency $\nu_0$, as at this frequency the CIB SED does not depend on $\beta$ and thus the derivative exactly vanishes\footnote{When we deproject components in a constrained NILC, the normalization of their SED does not matter (when we preserve a component in the NILC, its normalization is important in order to interpret correctly the units of the final map). In the case of the SED of the CIB, $\nu_0$ appears as a normalization and can be divided out. The same is not true for the first moment with respect to $\beta$, as $\nu_0$ does not appear multiplicatively.}. In all cases, we use a pivot frequency of $\nu_0=353 $ GHz.

We refer the interested reader to Ref.~\cite{Paper2} for details of how this moment deprojection technique stabilizes the measurement of $C_\ell^{y\kappa}$ even for different choices of the CIB SED parameters $(\beta, T_{\mathrm{CIB}}^{\mathrm{eff}})$.

\subsection{Deprojection of the CMB}

The CMB, whose SED is known (nearly) perfectly, can be deprojected in a constrained ILC tSZ map with no uncertainty or need to consider additional moments. This can be useful for large-scale cross-correlations of the tSZ signal with LSS, as the large-scale CMB contains contributions sourced by LSS arising from the integrated Sachs-Wolfe (ISW) effect~\cite{1967ApJ...147...73S}.  This ISW-LSS correlation (e.g.,~\cite{2011JCAP...03..018L,2016PhRvD..94f3519C,2018PhRvD..98h3542H}) can bias such measurements, including the tSZ -- CMB lensing cross-correlation. However, we do not have enough frequency channels to deproject more than three components at small scales (unless 857 GHz information is included); see Table~\ref{tab:beam_criterion_Scales}. Thus we elect to only deproject the CMB from the first five needlet scales in some cases, denoted ``CMB${}^5$''.  As the ISW is a large-scale effect, this should mitigate essentially all of the LSS bias due to the ISW contribution.

\begin{figure}
\includegraphics[width=0.5\textwidth]{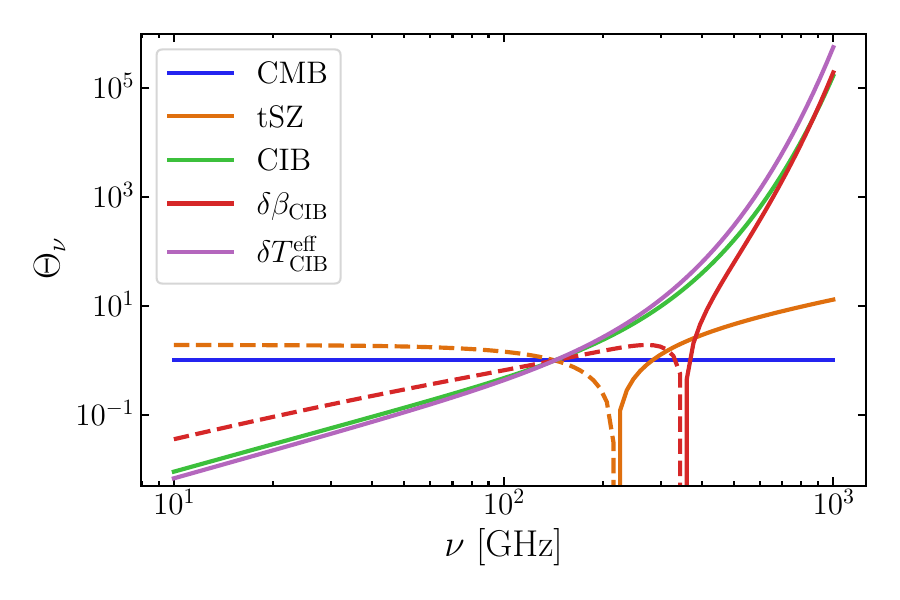}
\caption{The SEDs of various components of interest: the tSZ effect (orange), which we want to isolate; the CIB (green), which we wish to deproject; and the first moments of the CIB with respect to $\beta$ (red) and $T_{\mathrm{CIB}}^{\mathrm{eff}}$, which we also wish to deproject. The SEDs are shown in thermodynamic temperature units in which the blackbody CMB is constant, as indicated by the CMB SED in blue. Due to the logarithmic $y$-axis, we show negative values with dashed lines. We normalize all of the SEDs so that their absolute value is unity at 143 GHz.  In all cases we use the best-fit CIB SED parameters of $\beta = \bestfitbetastandard$, $T_{\mathrm{CIB}}^{\mathrm{eff}} = \bestfitTstandard \,\, \mathrm{K}$. }\label{fig:SEDs}
\end{figure}

The (normalized) SEDs of various components of interest are plotted in Figure~\ref{fig:SEDs}.

\section{Compton-$y$ map validation and comparison to \textit{Planck} map}\label{sec:comp_planck}

In this section we present and validate our Compton-$y$ maps, estimate their power spectra, and compare them to the official \textit{Planck} 2015 tSZ NILC map. We explicitly compute the auto-power spectra and 1-D histograms of our maps and compare them to those of the official \textit{Planck} analysis (on the same region of sky) in Section~\ref{sec:autospectra}. In Section~\ref{sec:deprojections} we quantify the variance increase resulting from our various deprojections, by comparing the power spectra of our undeprojected and deprojected maps; we also illustrate the effects of varying the frequency coverage in the NILC. Finally, in Section~\ref{sec:CIB_contamination}, we compare the level of CIB contamination in our maps to that of the official \textit{Planck} release, and also present the results of deprojecting the CIB. 
We perform all power spectrum calculations throughout with \texttt{NaMaster}~\cite{2019MNRAS.484.4127A}. 

Note that the 2015 \textit{Planck} map was made with the CMB deprojected, and so all direct comparisons between our map and the \textit{Planck} map are done with our CMB-deprojected map.

\subsection{Map images}

\begin{figure*}
\includegraphics[width=0.32\textwidth]{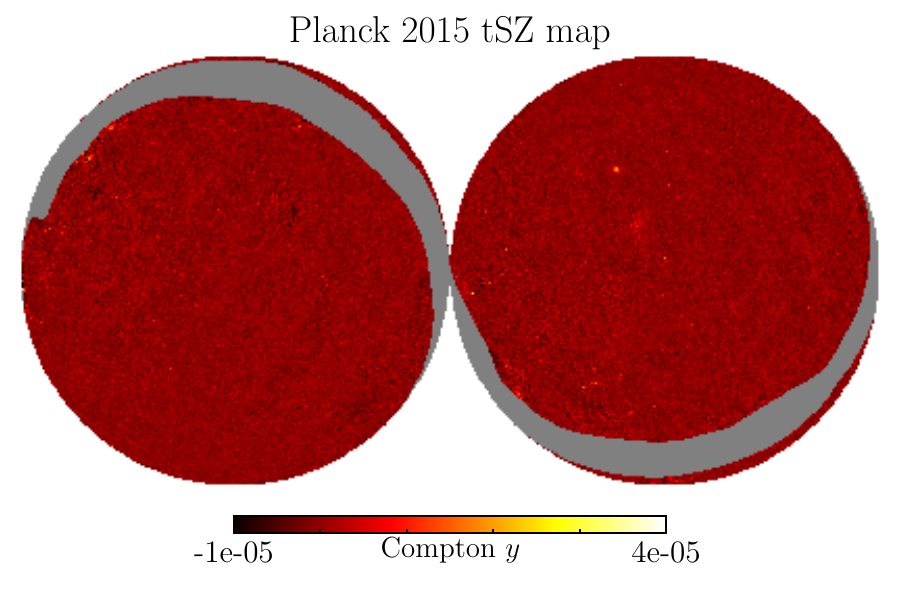}
\includegraphics[width=0.32\textwidth]{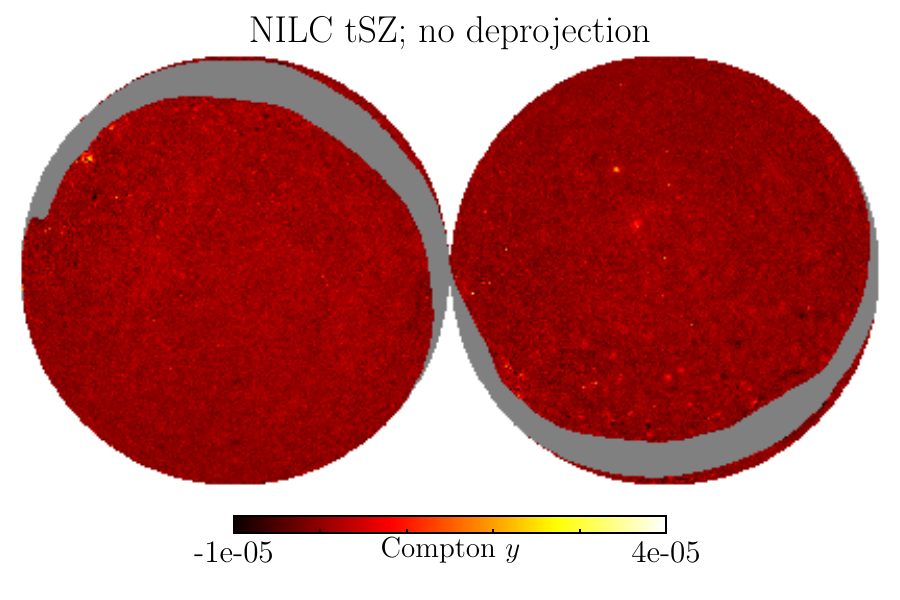}\\
\includegraphics[width=0.32\textwidth]{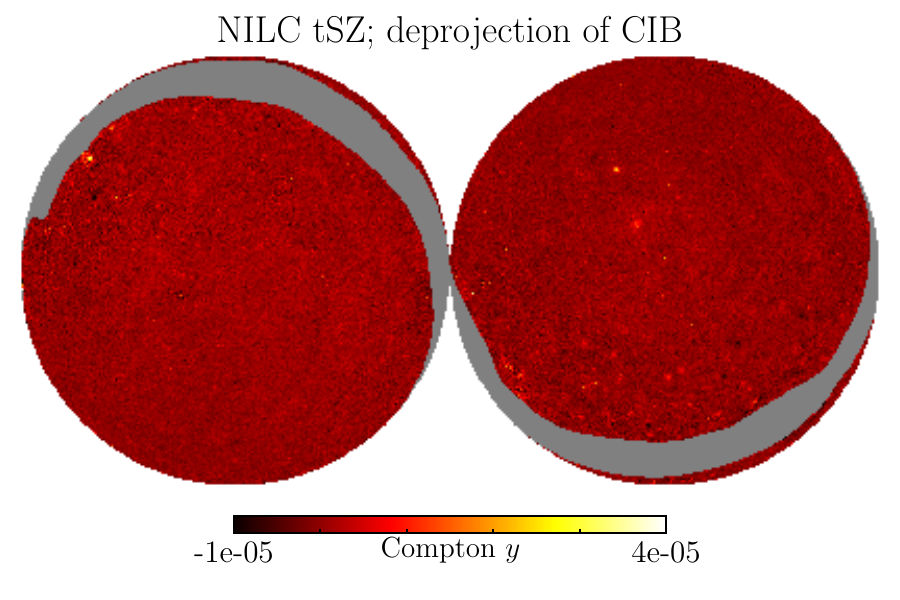}
\includegraphics[width=0.32\textwidth]{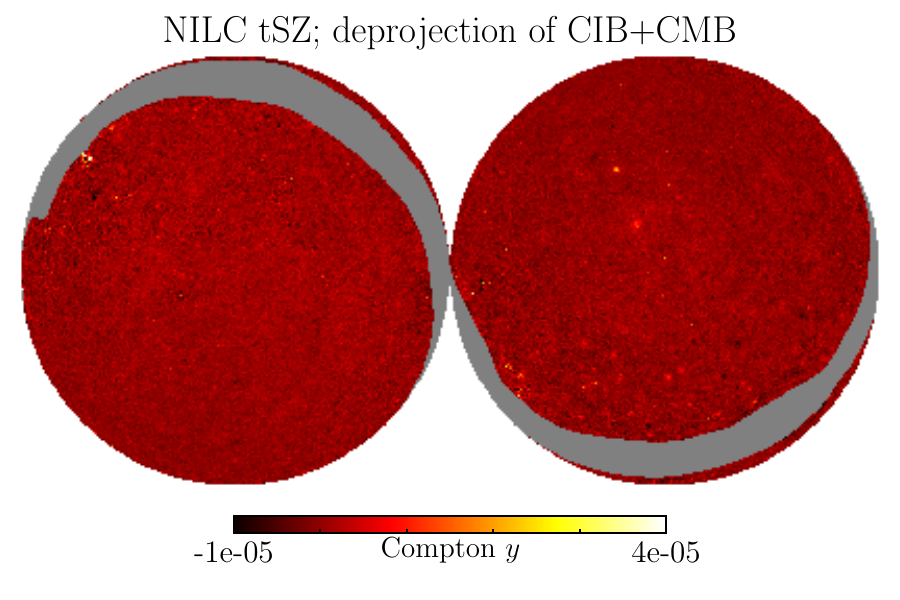}\\
\includegraphics[width=0.32\textwidth]{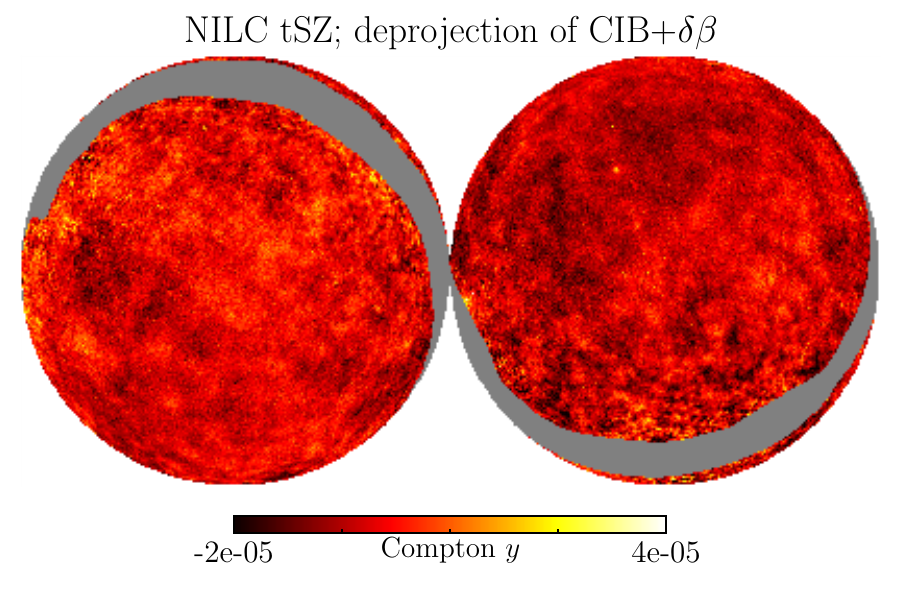}
\includegraphics[width=0.32\textwidth]{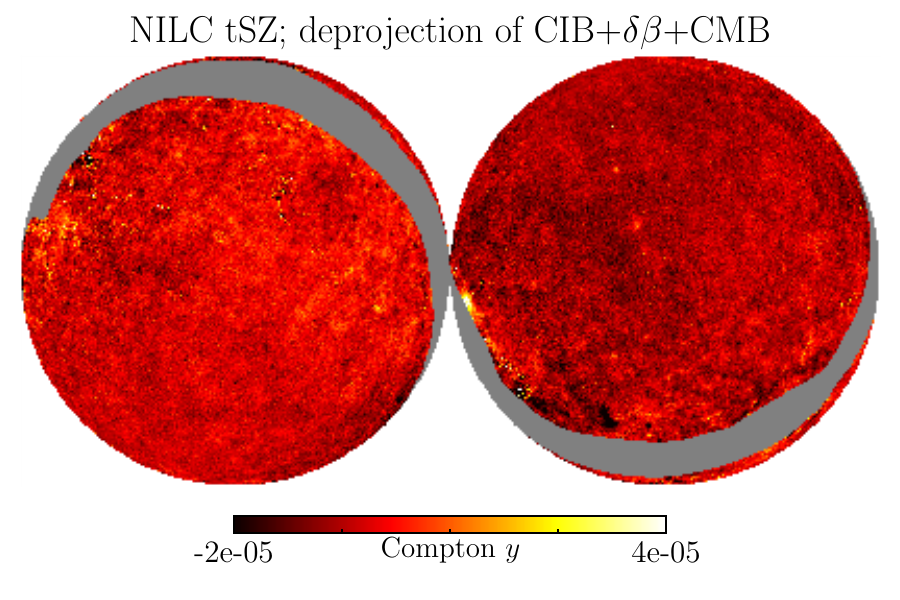}\\
\includegraphics[width=0.32\textwidth]{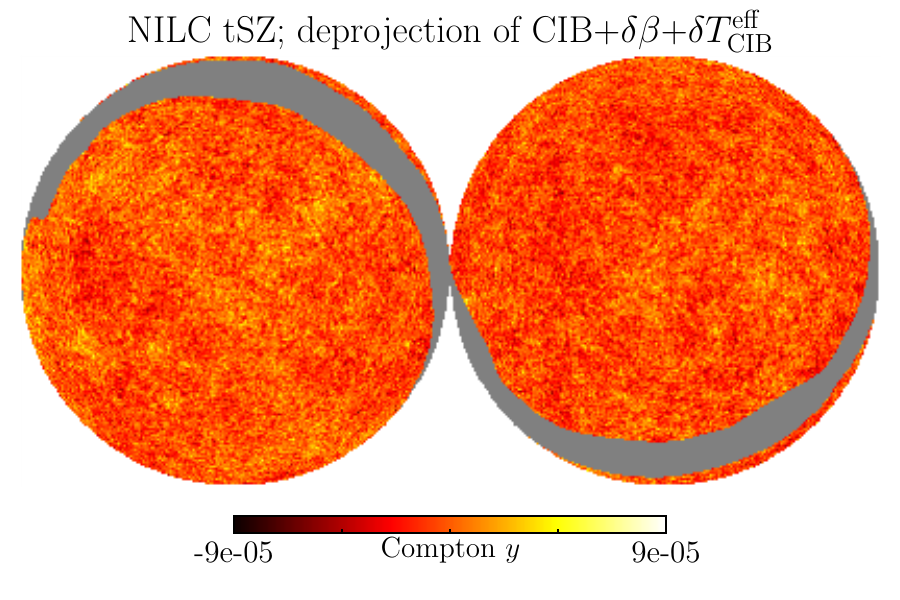}
\includegraphics[width=0.32\textwidth]{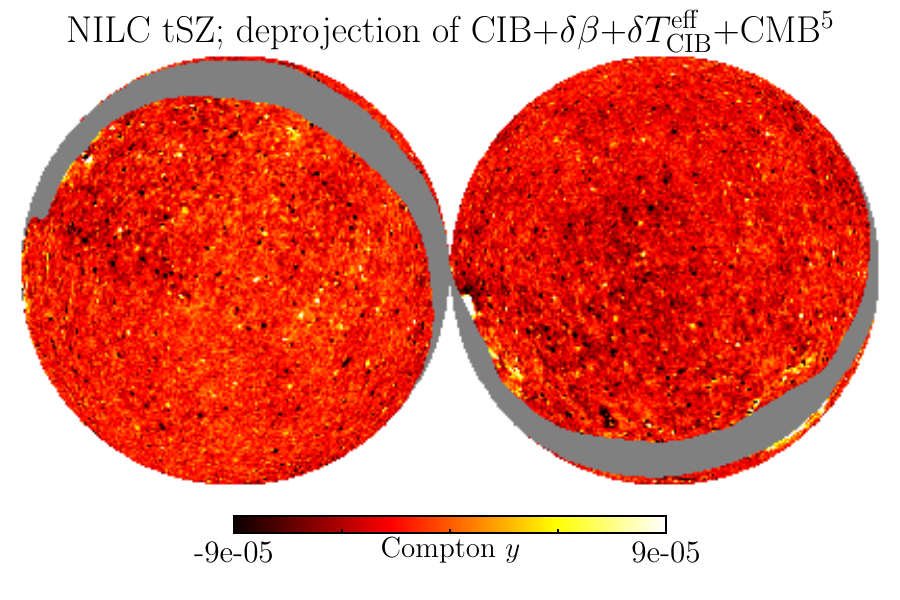}

\caption{ Our needlet ILC maps, visualized in orthographic projection in equatorial coordinates.  In all cases, the northern hemisphere is on the right and the southern on the left. We also show, on the top left, the official \textit{Planck} 2015 NILC tSZ map~\cite{2016A&A...594A..22P}. In all cases we have masked out the Galaxy with the \textit{Planck} Galactic plane mask, which covers 20\% of the sky. Note that for these visualizations, we show Gaussian-beam-convolved maps, which have FWHM $= 10^\prime$. We note that the \Planck $y$-map has deprojected the CMB; we do not show our CMB-deprojected map here but it is indistinguishable by eye from the no-deprojection, CIB-deprojected, and CIB+CMB-deprojected maps. Indeed, it is only when we deproject $\delta\beta$ that adding the CMB deprojection makes a visible difference (as seen in the third row). Note the increased ranges on the color bars in the bottom four plots, due to their significantly increased variances.}\label{fig:visualize_maps}
\end{figure*}

We show our resulting tSZ maps in Figure~\ref{fig:visualize_maps}, including various deprojection choices and also showing the official \textit{Planck} 2015 NILC tSZ map~\cite{2016A&A...594A..22P} for comparison. For the cases when just the CIB or the CIB+CMB are deprojected, there is no significant change to the maps by eye. However, when we deproject $\delta\beta$, the map becomes visibly noisier; we see that when we deproject both $\delta\beta$ and the CMB the map is even noisier, indicating that there is significant CMB contamination in the CIB+$\delta\beta$-deprojected map. Finally, the maps with both moments of the CIB deprojected have significantly higher variance, and it becomes difficult to pick out structures such as the Coma cluster, which appears as a bright point in the northern hemisphere (right hand side of the maps) by eye. In the legend of Figure~\ref{fig:visualize_maps}, and later, CMB$^5$-deprojection refers to the deprojection of the CMB on only the first five needlet scales.

\subsection{Comparison with \textit{Planck} map: histograms and power spectra}\label{sec:autospectra}

To compare directly with the \textit{Planck} $y$-map, we consider histograms of the Compton-$y$ values in the maps and also compute their auto-power spectra (for our various deprojection choices) on the area of sky defined by the \textit{Planck} $y$-map analysis mask, which is provided with the \textit{Planck} $y$-map\footnote{It can be found at \url{http://pla.esac.esa.int/pla/aio/product-action?MAP.MAP_ID=COM_CompMap_Compton-SZMap-masks_2048_R2.01.fits}, with \texttt{field=1}.}. This mask, which is apodized, covers 50\% of the sky. We multiply this mask by the point source mask, defined by the union of the \textit{Planck} HFI and LFI point source masks as described in Ref.~\cite{2016A&A...594A..22P}.  When we estimate the power spectra, we apodize the point source mask with an apodization scale of $30^\prime$ before multiplying it with the \textit{Planck} analysis mask (for the histograms, we use an unapodized point source mask). The resulting apodized mask allows for analysis with an effective $f_{\mathrm{sky}} = 0.45$\footnote{Note that, at some level, the point source mask is correlated with the tSZ field itself; however, any biases resulting from this correlation (e.g.,~\cite{2021PhRvD.103d3535F,2022PhRvD.106b3525L,2023PhRvD.107h3521S}) are expected to be negligible at \emph{Planck} sensitivity.}.

\subsubsection{Histograms}

\begin{figure}[t!]
\includegraphics[width=\textwidth]{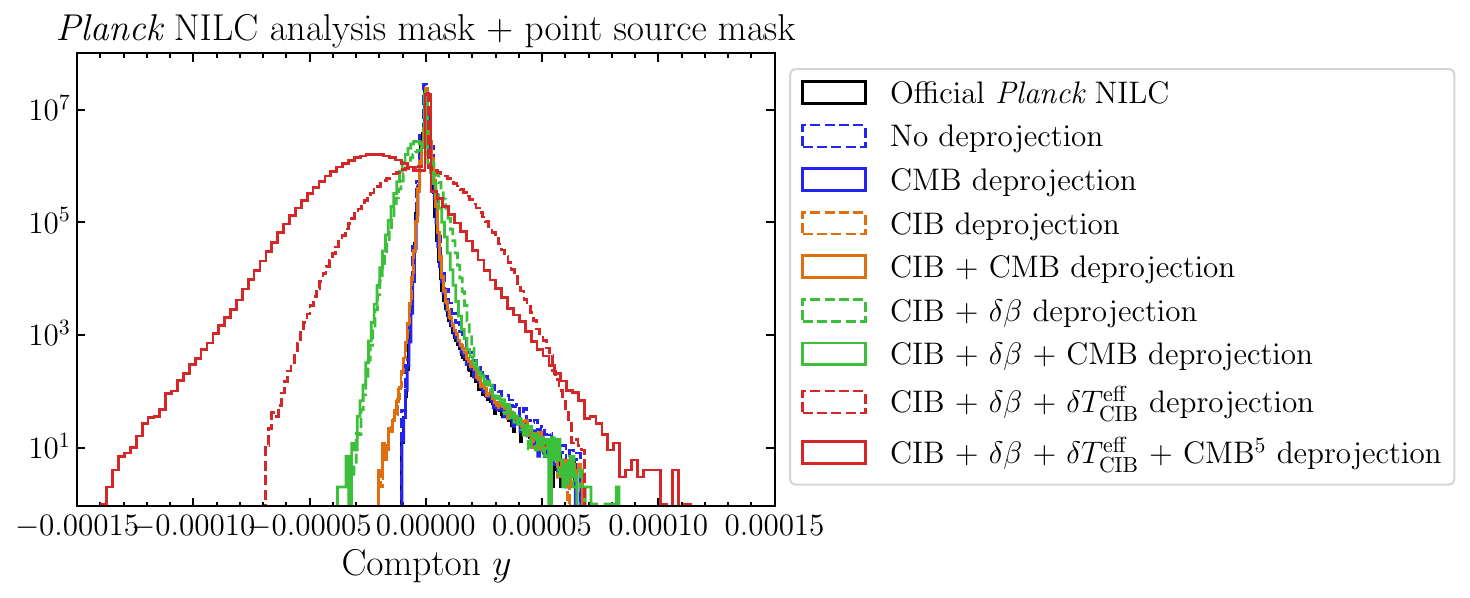}
\caption{Histograms of the Compton-$y$ values of our tSZ maps (for various deprojection choices as labeled), as compared to that of the official \emph{Planck} 2015 tSZ map. These histograms are estimates of the one-point probability density function (PDF) of the tSZ effect (which itself is a sensitive cosmological probe~\cite{2014arXiv1411.8004H,2019PhRvD..99j3511T}). We show the histograms of the maps convolved with a FWHM $= 10^\prime$ beam, and apply the \textit{Planck} NILC analysis mask and the point source mask described in the text. The positive tails in the histograms indicate the presence of tSZ signal, with the increased noise in the CIB-deprojected and CIB+$\delta\beta$ maps evident as an increased spread around the peak at $y=0$. The histogram of our no-deprojection and our CMB-deprojected maps agree very well with each other and with that of the official \textit{Planck} map, demonstrating how small the signal-to-noise penalty is for deprojecting the CMB; the CIB-deprojected versions are slightly noisier (note that the CIB+CMB-deprojected version is not meaningfully noisier than the CIB-deprojected version, with the orange dashed and solid lines lying on top of each other). Once we deproject the moments of the CIB, there is a larger increase in noise, although for the $\delta\beta$-deprojected version the non-Gaussian positive tail is still visible. When we deproject both moments, the contribution from foregrounds is is larger than this tail and it cannot be seen by eye; we also see that for the case when both moments are deprojected there is a significant noise penalty when we also deproject the CMB (note that we do not deproject the CMB on all scales, as we do not have enough frequency coverage; instead, we just do so on large scales, in particular in the first five needlet scales, referred to as ``CMB$^5$- deprojection'' in the legend). }\label{fig:tsz_histogram_pdfs}
\end{figure}

We show the histograms of the Compton-$y$ pixel values in Figure~\ref{fig:tsz_histogram_pdfs}, for our maps with no deprojection (standard ILC, blue), fiducial CIB modified blackbody SED deprojected (orange), and additionally with the first moment of the CIB SED with respect to $\beta$ deprojected (green) and with both moments deprojected (red); we include all versions both with and without the CMB deprojected, with the CMB-deprojected versions shown in solid lines compared to the dashed lines for the cases when we do not deproject the CMB (we only deproject the CMB on the first five needlet scales in the case where we deproject both moments of the CIB due to the insufficient frequency coverage on small scales to deproject more components).  We also show the histogram for the official \emph{Planck} 2015 tSZ map in black.  The Compton-$y$ signal is highly non-Gaussian and the strong positive tail is indicative of the presence of tSZ signal due to groups and clusters, which we see in all of the maps except for the case when we deproject both moments of the CIB; in this case, the large foreground contribution is larger than this signal. We note that there is more noise in the CIB-deprojected maps, and significantly more noise in the moments-deprojected maps, as indicated by the broadening of the histograms.  This is an inevitable consequence of deprojecting more components in a constrained ILC.

We note that the histograms of our no-deprojection (standard ILC) and CMB-deprojected $y$-maps agree very well with that of the \textit{Planck} map.

\subsubsection{Auto-power spectra}

\begin{figure}[t!]
\includegraphics[width=0.8\textwidth]{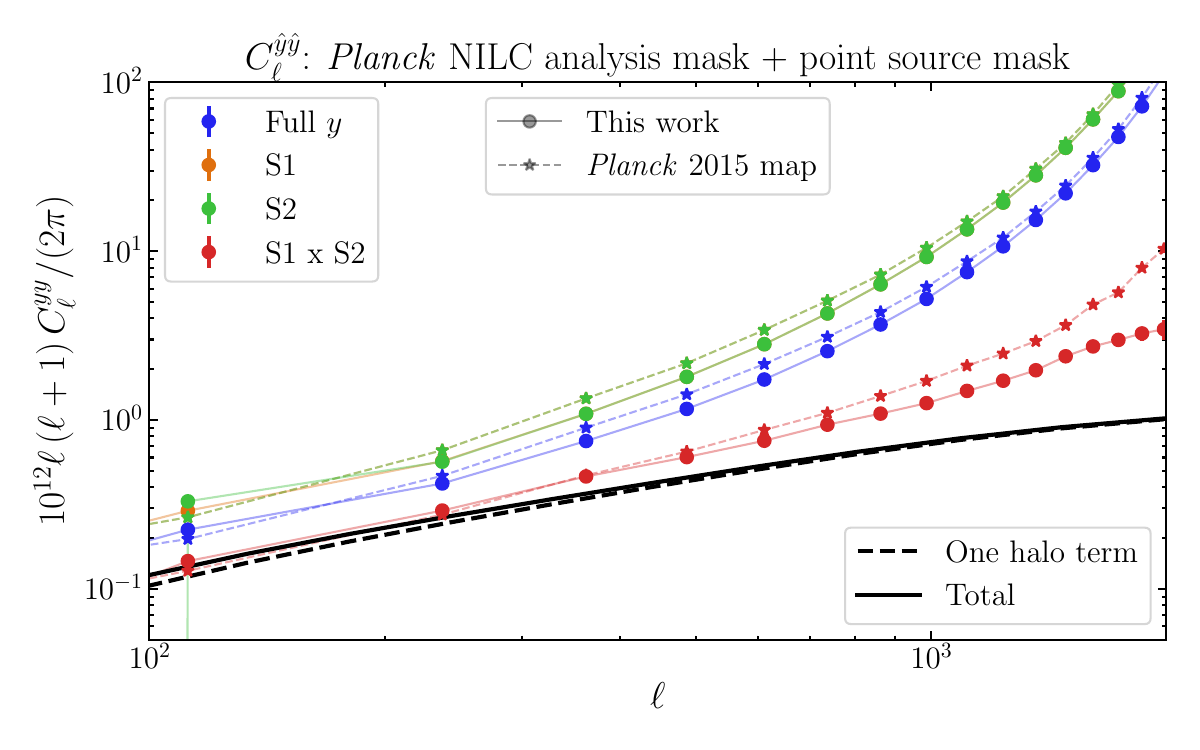}
\caption{The auto-power spectrum of our Compton-$y$ maps, estimated by taking the auto-power of the full-mission $y$-map (blue points) and also of the split maps (orange and green points, which nearly coincide), and the cross-power spectrum of the split maps (red points).  We show the power spectra estimated both from our maps (circles) and from the 2015 \textit{Planck} tSZ map (stars). Our maps show a clear reduction in noise power, as assessed from the auto-spectra, and also a reduction in the residual foreground power, as assessed from the cross-spectra, compared to the 2015 \textit{Planck} release.  For reference, we also show a theoretical prediction of the tSZ power spectrum in black (total signal in solid, one-halo term only in dashed). }\label{fig:power_spectra_planckmask_withsplits}
\end{figure}

We measure the power spectra of our $y$-maps to further characterize their properties.  We also construct NILC $y$-maps with, and measure the cross-power spectra of, two independent split maps which have independent noise (and therefore no noise bias in the power spectrum). We show the results in Figure~\ref{fig:power_spectra_planckmask_withsplits}. We also calculate the auto-power spectrum of the \textit{Planck} $y$-map on the same sky area. In all cases, we calculate the power spectra using \texttt{NaMaster} to decouple the mask mode-coupling matrix.

In Figure~\ref{fig:power_spectra_planckmask_withsplits}, and in all subsequent plots with auto-power spectra, we include a theoretical calculation of the tSZ power spectrum, calculated using \texttt{class\_sz}\footnote{\url{https://github.com/borisbolliet/class_sz}}~\cite{2018MNRAS.477.4957B,2020MNRAS.497.1332B,2022arXiv220807847B}, which is an extension of the cosmological Boltzmann solver \texttt{class}\footnote{\url{https://lesgourg.github.io/class_public/class.html}}~\cite{2011JCAP...07..034B}. This signal is computed in the halo model using the pressure-mass relation of Ref.~\cite{2012ApJ...758...74B}.  We refer the reader to~\cite{Paper2} for a detailed discussion of the modeling of this signal.

In general, the power spectrum error bars can be estimated with the Gaussian expression for the covariance: 
\be
\mathbb{ C}( \hat C^{\alpha\beta }_\ell,\hat C^{\gamma\delta }_{\ell^\prime})= \frac{\delta_{\ell \ell^\prime}}{(2\ell+1 ) f_{\mathrm{sky}}}\lb\lb C_\ell^{\alpha\gamma} + N_\ell^{\alpha\gamma}\rb\lb C_\ell^{\beta\delta} + N_\ell^{\beta\delta}\rb+\lb C_\ell^{\alpha\delta} + N_\ell^{\alpha\delta}\rb\lb C_\ell^{\beta\gamma} + N_\ell^{\beta\gamma}\rb\rb \,,\label{general_cov_matrix_cell}
\ee
where $C_\ell^{\alpha\beta}$ and $N_\ell^{\alpha\beta}$ indicate the signal and noise including all sources of foregrounds; in practice, we replace these with the measured power $\hat C_\ell^{\alpha\gamma}$ such that
\be
\sigma^2 \lb\hat {C}^{yy }_\ell\rb = \frac{2\, \lb \hat C_\ell^{yy}\rb^2}{(2\ell+1 ) f_{\mathrm{sky}}} \,.\label{errorbars_cyy}
\ee
We do not calculate Eq.~\eqref{general_cov_matrix_cell} directly, but instead use \texttt{NaMaster}, which computes the Gaussian covariance accounting properly for the decoupling of the mask. Note that this requires us to have an estimate of $C_\ell+N_\ell$ at every $\ell$, which we achieve by unbinning our measured $\hat C_\ell$.

The auto-power spectrum of our standard ILC $y$-map is slightly lower ($\approx 10-20\%$) on small scales than that of the \textit{Planck} 2015 $y$-map, presumably due to the lower instrumental noise in the PR4 \NPIPE\ data compared to PR3. 
In addition, we also see a significant decrease in the cross-power spectrum of our split maps compared to that of the \textit{Planck} 2015 $y$-map splits, except for on the largest scales.  This is presumably due to the improved foreground cleaning in our map, resulting from the lower noise in the \texttt{NPIPE} maps. 

We defer a full analysis of the tSZ auto-power spectrum from our maps to future work, as this will require careful understanding of the propagation of foreground contaminants, likely using simulations (as in Ref.~\cite{2016A&A...594A..22P}).

\subsection{Effect of various deprojections and varying frequency coverage}\label{sec:deprojections}

\begin{figure}[t!]
\includegraphics[width=\textwidth]{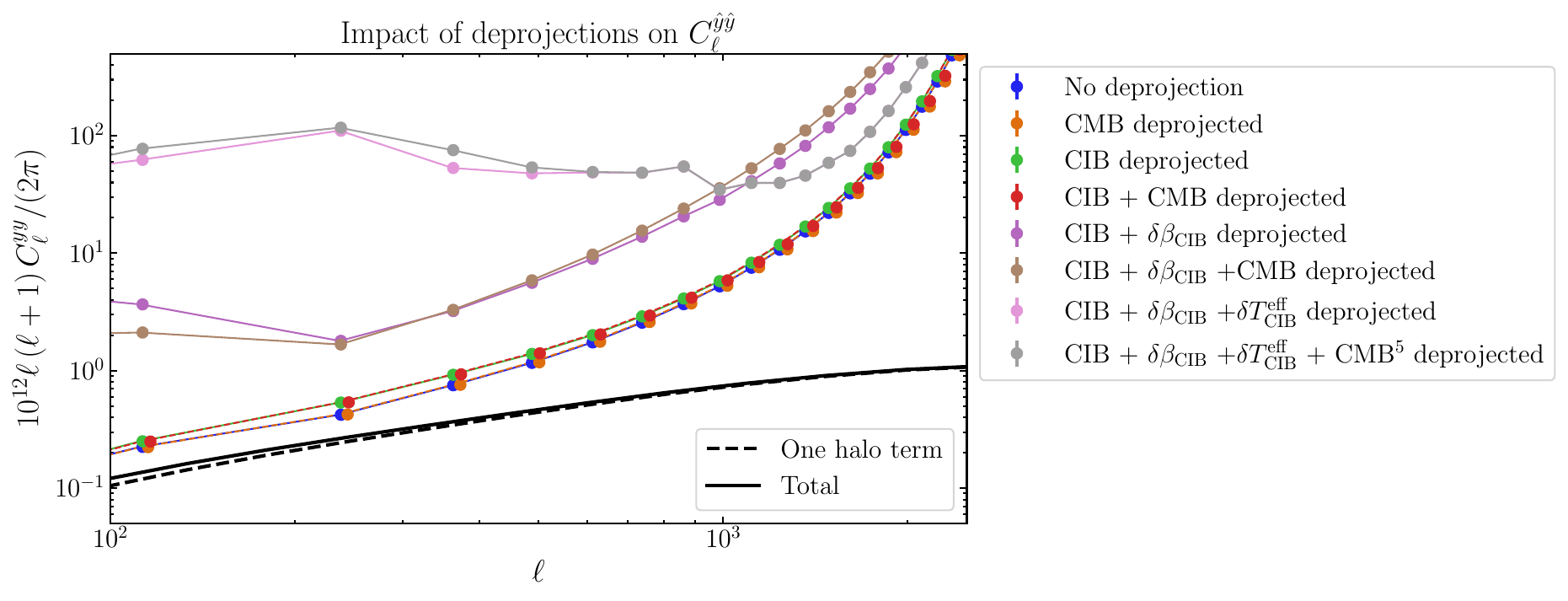}
\caption{Auto-power spectra of our tSZ maps for different deprojection choices, on the region of sky defined by the \textit{Planck} analysis mask and our point source mask. Note that, to aid in distinguishability, we have offset horizontally the orange and red CMB-deprojected and CIB+CMB-deprojected points slightly, as otherwise they would fall almost exactly on the blue and green points respectively. We have also connected these points with a dashed instead of a solid line so that the lines underneath are visible. We see that the first CMB deprojection is almost free in terms of its noise penalty; for CIB deprojection, in contrast, there is a slight noise penalty, corresponding to an increase of $\approx 10\%$ in the power (although an additional CMB deprojection here continues to increase the power negligibly).  Deprojecting the first moments of the CIB, however, incurs a much larger noise penalty, with the power increasing by a factor of $\approx 5$ compared to the no-deprojection power when we include $\delta \beta$ in the deprojection.  Additionally, deprojecting the CMB is no longer ``free'' in this case, with the brown points slightly higher than the purple points. Finally, deprojecting both $\delta \beta$ and $\delta T_{\mathrm{CIB}}^{\mathrm{eff}}$ incurs a serious penalty, with the power increasing by orders of magnitude at low $\ell$, although surprisingly decreasing at high $\ell$ compared to the CIB$+\delta \beta$ deprojection. 
As there are not enough frequency channels to simultaneously deproject all of these components \textit{and} the CMB at high resolution, we only do so for the first five needlet scales (grey points); this again incurs a small penalty compared to the case without deprojecting the CMB. }\label{fig:auto_power_deprojections}
\end{figure}

\begin{figure}[h!]
\includegraphics[width=\textwidth]{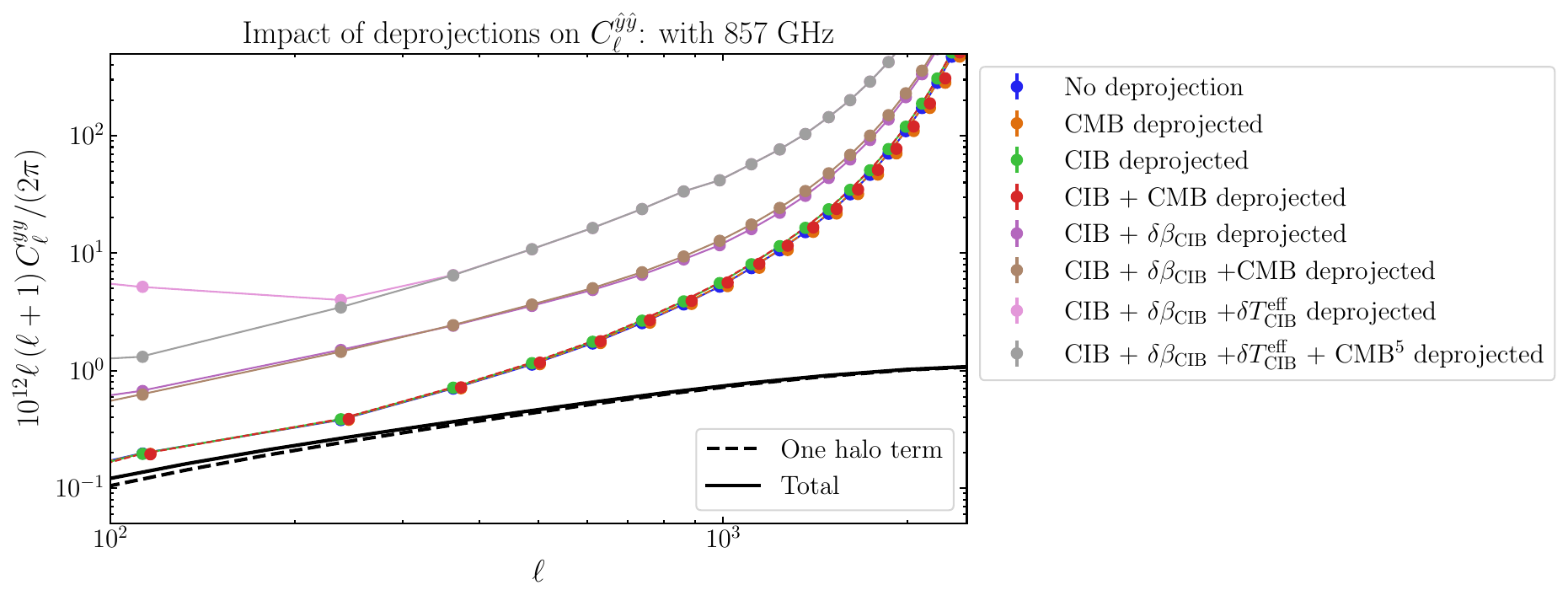}
\caption{The effect of different deprojections on a $y$-map constructed using a NILC that includes the 857 GHz channel. The noise penalty incurred when deprojecting many moments of the CIB is lower here than in the case when 857 GHz is not included in the NILC (see Figure~\ref{fig:auto_power_deprojections}). Again, we have slightly offset the CMB-deprojected (orange) points from the undeprojected (blue) points and the CIB+CMB-deprojected (red) points from the CIB-deprojected (green) points. }\label{fig:auto_power_deprojections_857}
\end{figure}

\begin{figure}[h!]
\includegraphics[width=\textwidth]{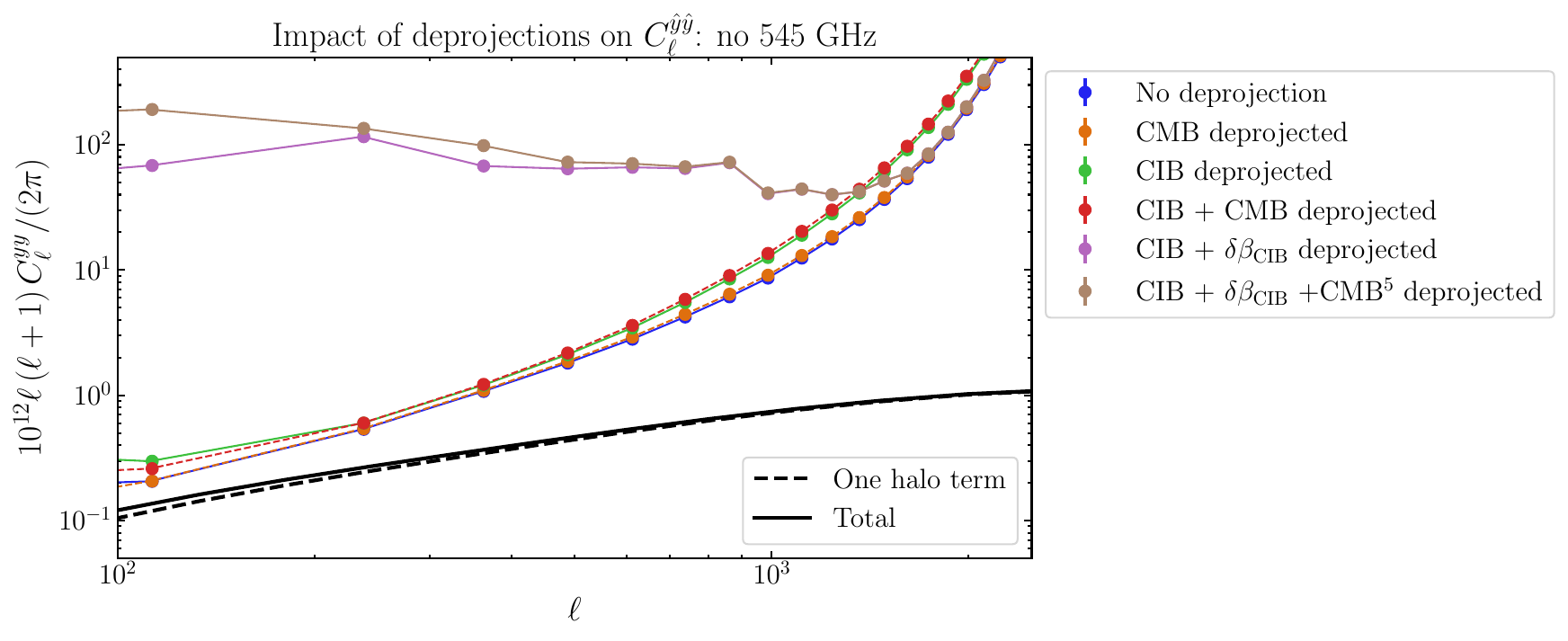}
\caption{The effect of different deprojections on a $y$-map constructed using a NILC that excludes the 545 GHz channel (857 remains excluded as well). There are not enough frequency channels to simultaneously deproject the CIB along with $\delta \beta$ and $\delta T_{\mathrm{CIB}}^{\mathrm{eff}}$ in this case. Note that, in contrast to Figure~\ref{fig:auto_power_deprojections}, we have not offset the orange CMB-deprojected points or the red CIB+CMB-deprojected points as they are more distinguishable by eye here than they are in Figure~\ref{fig:auto_power_deprojections}, due to the slightly increased noise penalty when deprojecting the CMB.}\label{fig:auto_power_deprojections_no545}
\end{figure}

In Figure~\ref{fig:auto_power_deprojections}, we show how various contaminant deprojection choices used in our NILC analyses affect the final auto-power spectrum of the $y$-map. In all cases, when we deproject the CIB and its first moments, we use the best-fit CIB SED parameters from Section~\ref{subsec:CIBSED}, $\beta=\bestfitbetastandard$ and $T_{\mathrm{CIB}}^{\mathrm{eff}}=\bestfitTstandard \, \mathrm{K}$. We see that deprojecting the CIB incurs a small increase in noise power, but the CMB deprojection leaves the power essentially unaffected, as long as either no other components or the CIB alone is deprojected.  This is because the CMB is already the dominant astrophysical contaminant in the standard ILC, so the weights are quite similar regardless of whether it is deprojected.  However, deprojecting the moments of the CIB incurs much more serious noise penalties, particularly if multiple moments are deprojected.  Note that, for the case when we deproject both moments of the CIB, there are not enough frequency channels to also deproject the CMB on small scales, and so we only deproject the CMB in the first five needlet scales; we refer to this in the legends of Figures~\ref{fig:auto_power_deprojections},~\ref{fig:auto_power_deprojections_857}, and ~\ref{fig:auto_power_deprojections_no545} as ``CMB$^5$''.

Our fiducial $y$-map includes information from all \textit{Planck} frequency channels except for 857 GHz.  For comparison, we also construct maps including 857 GHz in the NILC, as well as dropping 545 GHz.  We note that the official \textit{Planck} analysis (as well as the analysis of Ref.~\cite{2023arXiv230510193C}) uses 857 GHz only at $\ell<300$, due to its uncertain calibration. In principle, one could co-add the $\ell<300$ modes of our with-857 GHz map with the $\ell>300$ modes of our no-857 GHz map to obtain such a map. 

The effects of including 857 GHz and dropping 545 GHz are shown in Figures~\ref{fig:auto_power_deprojections_857} and~\ref{fig:auto_power_deprojections_no545}, respectively. As expected, as there is more (less) data, including more (fewer) frequency channels results in lower (higher) power in the final $y$-map. However, we note that when more high-frequency channels are included, more information about the CIB SED is required in order to deproject it robustly, and it can be more difficult to robustly remove all CIB contamination; we discuss this issue in detail in our companion paper~\cite{Paper2}.

\subsection{CIB contamination}\label{sec:CIB_contamination}

We attempt to quantify the amount of residual CIB in our Compton-$y$ maps by directly cross-correlating them with maps of the CIB.  CIB maps are difficult to construct due to the difficulty of separating CIB emission from Galactic dust emission, which is dominant on large scales and has a very similar frequency dependence to the CIB.  We use the CIB maps of Ref.~\cite{2019ApJ...883...75L}, which were constructed by using HI data~\cite{2016A&A...594A.116H} as a tracer of the Galactic dust in order to clean the dust emission from the highest-frequency (353, 545, 857 GHz) \textit{Planck} channels.  The resulting maps of the CIB emission at these frequencies are very clean of Galactic contamination.

For this investigation, we use only the 857 GHz CIB map, as the maps were constructed with the PR3 \textit{Planck} data and thus do not have maps built from the same independent noise realizations as the \NPIPE\ maps.  Thus, we do not have two independent splits that are required to make a measurement without instrumental noise bias.  However, as we do not use 857 GHz in our default NILC $y$-map construction, we can make a measurement with the 857 GHz map that is immune to noise bias.   It would be interesting in the future to repeat or extend the work of Ref.~\cite{2019ApJ...883...75L} to the \NPIPE\ data in order to measure cross-correlations with the lower frequencies that do not contain noise bias.  An additional advantage of using the 857 GHz map for the tSZ-CIB cross-correlation is that this is the channel in which the intrinsic $y$ signal is most suppressed compared to the CIB.  For lower frequencies (particularly 353 GHz and below), we would have to consider the $\left<y^{\rm CIB-map} y^{\mathrm{NILC-map}}\right>$ contribution to the signal.

\begin{figure}
\includegraphics[width=\textwidth]{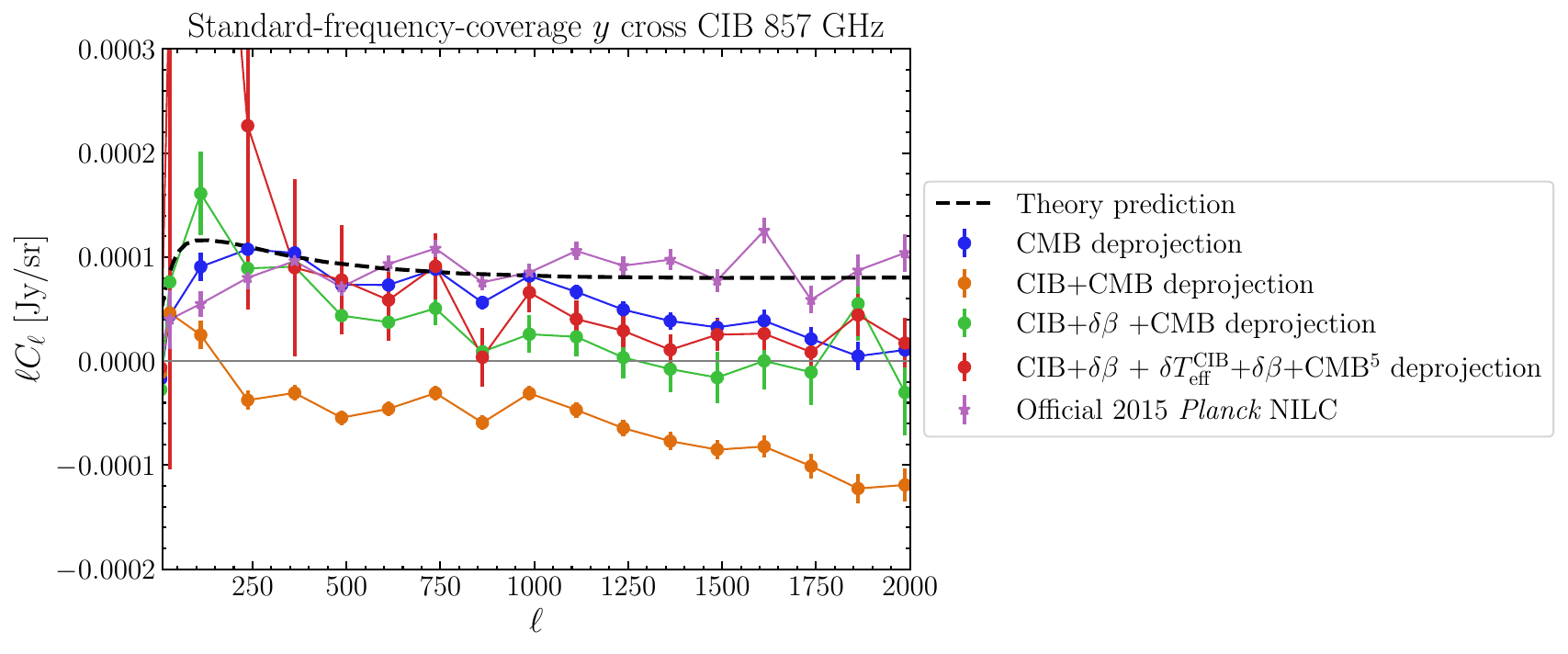}
\caption{The cross-correlation of our Compton-$y$ maps with the 857 GHz CIB map of Ref.~\cite{2019ApJ...883...75L}, for different deprojection choices, along with the measurement for the official \Planck\ NILC $y$-map (in black). We also show a prediction for the intrinsic $C_\ell^{y857}$ signal, calculated in \texttt{class\_sz} using the halo model (details in the main text). We note that the cross-correlation of our CMB-deprojected map with the CIB is significantly lower than that of the official \Planck\ map; a similar conclusion was found in Ref.~\cite{2023arXiv230510193C}. }
\label{fig:CIB_contamination}
\end{figure}

We show in Figure~\ref{fig:CIB_contamination} the cross-power spectra measured with the 857 GHz CIB map and our standard-frequency-coverage NILC $y$-maps for several deprojection choices.  We also include the measurement using the official \Planck\ map. We perform these measurements with the cleanest map provided by Ref.~\cite{2019ApJ...883...75L}, which covers 8.71\% of the sky.  We multiply the apodized mask provided by Ref.~\cite{2019ApJ...883...75L} with the mask used previously throughout this work, i.e., the official \Planck\ NILC mask combined with the point source masks.  The final sky covarage is 8.47\% in our tSZ-CIB cross-correlation measurement.

In Figure~\ref{fig:CIB_contamination}, it is clear that the CIB contamination in our $y$-maps is lower than that of the official \Planck\ $y$-map.  This is consistent with the results of Ref.~\cite{2023arXiv230510193C}.  It is encouraging that the lower instrumental noise in the \NPIPE\ maps has allowed for improved subtraction of the foregrounds in this case.

\begin{figure}
\includegraphics[width=0.75\textwidth]{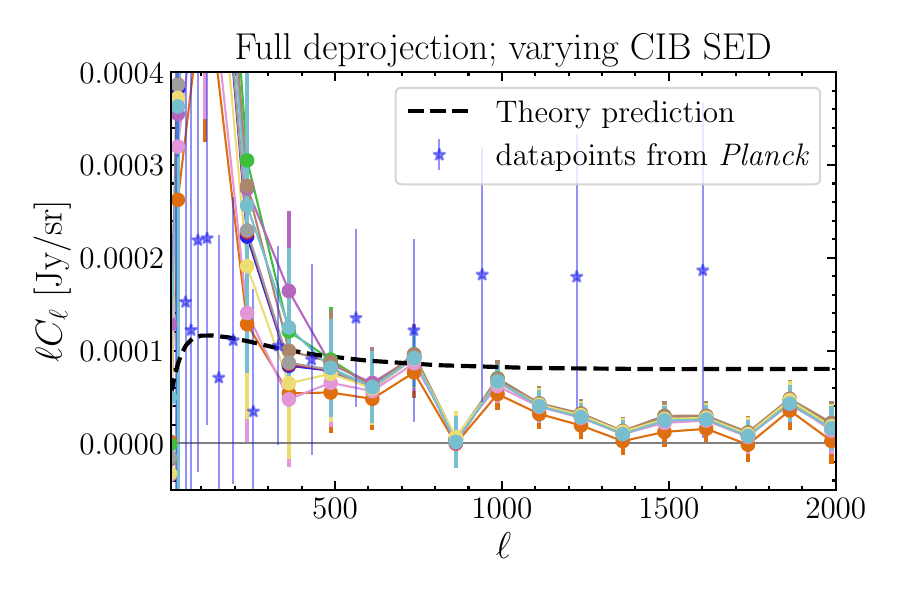}
\includegraphics[width=0.89\textwidth]{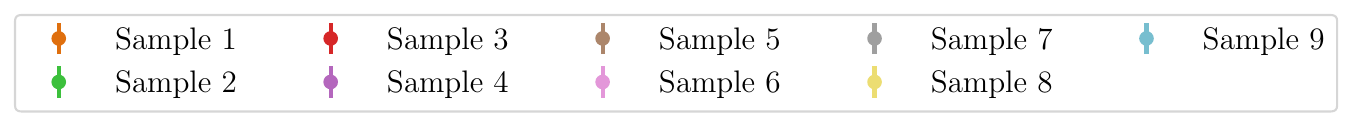}
\caption{The fully moment-deprojected (CIB+$\delta\beta$+$\delta T_{\mathrm{eff}}^{\mathrm{CIB}}$+CMB$^5$-deprojected) $y$-map cross-correlated with the 857 GHz CIB map from Ref.~\cite{2019ApJ...883...75L}, for various choices of the CIB SED consistent with the posterior in Figure~\ref{fig:posterior_betaT}.  We also include data points from the \textit{Planck} detection of this signal~\cite{2016A&A...594A..23P}, digitized from Figure 15 of that reference. The different-colored points correspond to different samples drawn from the posterior for $\beta,T$ shown in Figure~\ref{fig:posterior_betaT}. }
\label{fig:CIB_SED_moments}
\end{figure}

{In Figure~\ref{fig:CIB_contamination}, we also include a theoretical prediction for the intrinsic cross-power spectrum between the tSZ and the CIB emission at 857 GHz $C_\ell^{y {\rm 857}}$. We note that there is significant model uncertainty in this prediction. The methodology that we use follows the halo model formalism used to calculate the auto-power spectrum. While we defer to our companion paper~\cite{Paper2} an in-depth presentation of the halo model details, we note that the model in Figure~\ref{fig:CIB_contamination} (which was calculated with \texttt{class\_sz}) uses the pressure profiles of Ref.~\cite{2010ApJ...725...91B,2012ApJ...758...75B} to predict the tSZ signal along with the halo model of Ref.~\cite{2012MNRAS.421.2832S} with the parameter values of Ref.~\cite{2014A&A...571A..30P} to predict the CIB signal (see Ref.~\cite{2021PhRvD.103j3515M} for a detailed description of this incarnation of the CIB halo model). As the focus of this paper is not the CIB-tSZ cross-correlation, we do not consider a wider range of models (although note that a prediction for this signal was made in Ref.~\cite{2021A&A...645A..40M}, and it would be interesting to use this data to constrain this model).}
{We note that, if it is truly CIB-free, the fully moment-deprojected (CIB+$\delta\beta$+$\delta T_{\mathrm{eff}}^{\mathrm{CIB}}$+CMB$^5$) data points in Figure~\ref{fig:CIB_contamination} can be interpreted directly as a measurement of the intrinsic cross-correlation of Compton-$y$ and the CIB. Such a measurement was previously made with \Planck\ data in Ref.~\cite{2016A&A...594A..23P}. We defer to future work a comparison between our data and the measurements in this reference. Indeed, our work in Ref.~\cite{Paper2} relies on the assumption that the fully moment-deprojected $y$-map is indeed negligibly contaminated by CIB.  We note that the CIB+$\delta\beta$+$\delta T_{\mathrm{eff}}^{\mathrm{CIB}}$+CMB$^5$-deprojected points in Figure~\ref{fig:CIB_contamination} are indeed stable to variations in the assumed CIB SED, as shown in Figure~\ref{fig:CIB_SED_moments} (the same is not true of the CIB deprojection or the CIB+$\delta\beta$ deprojection). In Figure~\ref{fig:CIB_contamination}, we also include the data points as measured by Ref.~\cite{2016A&A...594A..23P} for $C_\ell^{y 857}$, which we have digitized from their Figure~15.\footnote{Note that their points are given in units of the Compton-$y$ parameter; we convert from Compton-$y$ to Jy/sr using the conversion factors appropriate for 857 GHz in Table 6 of Ref.~\cite{2014A&A...571A...9P}.} As such, we hope that these maps will make possible a fruitful joint CIB-tSZ analysis.}

\section{Discussion}\label{sec:conclusion}

In this work, we have presented \texttt{pyilc}, a fully public Python package to perform NILC on full-sky HEALPix maps. We hope that \texttt{pyilc} will be useful both for analyses of real data, to isolate maps of the CMB and tSZ signals and any other component whose spectral signature is known, as well as for analyses of \textit{simulations}, in particular in cases where one wants to test a signal-extraction pipeline on realistic data and to understand biases due to foregrounds.  We have validated that our implementation in \texttt{pyilc} produces maps that closely match the official \Planck tSZ maps (when making the same analysis choices), which were produced with a non-public pipeline.  Our code is fully available and easily extensible to consider other sky components and other deprojection options.

We have used \texttt{pyilc} to perform NILC on the \textit{Planck} \NPIPE\ data to isolate the (almost) full-sky Compton-$y$ signal in our Universe. We note that $y$-maps have previously been made with this data~\cite{2022MNRAS.509..300T,2023arXiv230510193C}, as well as in the official \Planck analysis~\cite{2014A&A...571A..21P,2016A&A...594A..22P}. We have made several improvements to the analysis, 
including a slightly different input-map-processing step where we mask the point sources such that they do not contribute to the calculation of the NILC weights; as point sources are always masked in any final analysis mask, this does not affect any science obtained from the maps, but improves their noise properties and usable sky fractions. Additionally, we provide $y$-maps where we have explicitly removed the CIB by deprojecting both a modified blackbody component and its first moments.  Our  {CMB-deprojected} tSZ map has $\approx 10$\% lower noise than that in the official \Planck\ 2015 tSZ map.

{We quantify the amount of CIB contamination in our maps by directly cross-correlating them with maps of the CIB. We find that our maps have less CIB contamination than the 2015 \Planck\ NILC map. We have also presented CIB-moment-deprojected $y$-maps, which we hope can be used to measure $\left<y-LSS\right>$ cross-correlations without CIB bias.}

We use these $y$-maps in our companion paper~\cite{Paper2} to detect the cross-correlation of the reconstructed CMB lensing signal and the tSZ signal, for only the second time~\cite{2014JCAP...02..030H}, and with significantly more robustness against CIB contamination~\cite{2015A&A...575L..11H}.  We hope that our techniques for removing the CIB from such a measurement will be useful for a wide range of future tSZ cross-correlation analyses, including the tSZ-CIB cross-correlation~\cite{2021A&A...645A..40M,2016A&A...594A..23P}.

We note that \Planck $y$-maps, while currently state-of-the-art, will shortly be surpassed in signal-to-noise ratio by upcoming data from ACT~\cite{2016JLTP..184..772H}, SPT-3G~\cite{2014SPIE.9153E..1PB}, and soon the Simons Observatory~\cite{2019JCAP...02..056A}.  However, the \Planck data will remain dominant in the Compton-$y$ reconstruction for the foreseeable future at $\ell \lesssim 1000$, as atmospheric noise becomes too large at low multipoles for ground-based experiments to surpass \Planck (in temperature) over this range --- see, e.g., Figure~36 of Ref.~\cite{2019JCAP...02..056A}.  A hybrid combination of the ground-based and space-based data will thus become the standard approach in the near future (as already demonstrated in Refs.~\cite{2020PhRvD.102b3534M,2019A&A...632A..47A}).

\section{Summary of Public Products}

We make various configurations of our Compton-$y$ maps publicly available at  \url{https://users.flatironinstitute.org/~fmccarthy/ymaps_PR4_McCH23/}. We also release our full NILC pipeline at \url{https://github.com/jcolinhill/pyilc/}, which takes as input a set of single-frequency maps and executes the NILC algorithms described in this paper; we include as explanatory files several input files we used to construct our NILC $y$-maps from the \textit{Planck} data products. We note that the input is not the raw single-frequency \NPIPE\ maps but instead the monopole- and dipole-subtracted, inpainted \NPIPE\ maps described earlier in this work. We also release our inpainting code in the same GitHub repository as {\tt pyilc},  along with our inpainting mask and several analysis masks, which are also at \url{https://sdsc-users.flatironinstitute.org/~fmccarthy/ymaps_PR4_McCH23/}.

\begin{acknowledgements}
We are very grateful to Will Coulton for many useful discussions including comparisons of ILC pipelines. We are also very grateful to Boris Bolliet for discussions about the halo model and \texttt{class\_sz}. We also thank Shivam Pandey for a useful consistency check of our $y$-maps. We thank Jens Chluba for discussions about the moment expansion method. We additionally thank Aleksandra Kusiak, Mathew Madhavacheril, Mathieu Remazeilles, Blake Sherwin, David Spergel, and Kristen Surrao for useful conversations. We thank the Scientific Computing Core staff at the Flatiron Institute for computational support. The Flatiron Institute is supported by the Simons Foundation. JCH acknowledges support from NSF grant AST-2108536, NASA grants 21-ATP21-0129 and 22-ADAP22-0145, DOE grant DE-SC00233966, the Sloan Foundation, and the Simons Foundation.

We perform all map manipulations using the Python package \texttt{healpy}~\cite{Zonca2019}, an implementation of \texttt{HEALpix}~\cite{2005ApJ...622..759G}.

\end{acknowledgements}

\bibliography{references}

\appendix

\end{document}